\documentclass[a4paper,twocolumn,11pt,supersciiptaddress,accepted=2024-08-20]{quantumarticle}
 \pdfoutput=1
\usepackage{ORI_Group_style}
\usepackage{blindtext}
\usepackage[dvipsnames]{xcolor}
\usepackage{array}
\usepackage{subfigure}
\usepackage{xurl}
\usepackage{float}
\usepackage{tabularx}
\usepackage[normalem]{ulem}
\usepackage[numbers]{natbib}

\newcommand{\Lv}{\mathcal{L}}
\newcommand{\Py}{\mathcal{P}}
\newcommand{\Qy}{\mathcal{Q}}
\newcommand{\Id}{\mathbb{1}}
\newcommand{\Idy}{\mathfrak{I}}

\hypersetup{colorlinks,linkcolor=red,citecolor=blue,urlcolor=blue}

\begin{document}

\title{Tutorial: projector approach to master equations for open quantum systems}

\author{C. Gonzalez-Ballestero}
\email{c.gonzalez-ballestero@uibk.ac.at}
\affiliation{Institute for Theoretical Physics, Vienna University of Technology (TU Wien), 1040
Vienna, Austria}

\begin{abstract}
Most quantum theorists are familiar with different ways of describing the effective quantum dynamics of a system coupled to external degrees of freedom, such as the Born-Markov master equation or the adiabatic elimination. Understanding the deep connection between these -- sometimes apparently  unrelated -- methods can be a powerful tool, allowing us to derive effective dynamics in unconventional systems or regimes.
This tutorial aims at providing quantum theorists across multiple fields (e.g., quantum and atom optics, optomechanics, or hybrid quantum systems) with a self-contained practical toolbox to derive effective quantum dynamics, applicable to systems ranging from $N-$level emitters to mechanical resonators. First, we summarize the projector approach to open quantum systems and the derivation of the fundamental Nakajima-Zwanzig equation. Then, we show how three common effective equations, namely the Brownian master equation, the Born-Markov master equation, and the adiabatic elimination used in atom and molecular optics, can be derived from different perturbative expansions of the Nakajima-Zwanzig equation. We also solve in detail four specific examples using this formalism, namely a harmonic oscillator subject to displacement noise, the effective equations of a mechanical resonator cooled by an optical cavity, the Purcell effect for a qubit coupled to an optical cavity, and the adiabatic elimination in a Lambda system.
\end{abstract}

%\pacs{Insert PACS}

\maketitle

 {
   \hypersetup{linkcolor=black}
   \tableofcontents
 }

\section{Introduction}

A core tool in theoretical quantum physics is the ability to derive, starting from the dynamical equation of a system, a consistent quantum dynamical equation only for a subsystem of interest~\cite{tutorial1,tutorial3,tutorial4,tutorial5,tutorial6}. 
The specific procedure, as well as the resulting effective equations, depends on the regime of operation and on the relations between system parameters. Some particularly useful and powerful techniques are, for instance, ``adiabatic elimination'', which can be used to trace out highly detuned levels of a lossless $N-$level system~\cite{BrionJPhysA2007,CiracPRL1997}, or the ``Born-Markov master equation'' which is typically employed to describe open systems coupled to a rapidly decaying bath~\cite{CiracPRA1992,DuanNature2001,WilsonRaeNJP2008}.
These and other examples are the quantum extension of well-known methods to eliminate fast variables in classical mechanics~\cite{VanKampen,VANKAMPEN1974215}, and have made the derivation of effective dynamical equations a core tool of modern quantum theory~\cite{BreuerPetruccione,weiss2012quantum}.

Many researchers are familiar with the above methods only via their common use in specific quantum physics problems. For instance, the quantum optical master equation describing qubit spontaneous decay into the free-space electromagnetic field is a Born-Markov master equation~\cite{CohenAtomPhotonInteractions}, whereas the elimination of far-detuned levels in few-level systems is a famous case of adiabatic elimination in quantum optics~\cite{BrionJPhysA2007}. 
Although needed to efficiently study paradigmatic systems, this ``case-by-case'' approach can make it difficult to derive effective equations for systems where conventional approaches are not applicable.
A solution to this problem is offered by a more general formulation of open quantum systems in terms of projection superoperators in density matrix space, which allows to derive a single general equation describing the quantum dynamics of subsystems, the Nakajima-Zwanzig equation~\cite{BreuerPetruccione,weiss2012quantum,GardinerZollerQNoise,toda2012statistical,Haake1973}. Any effective description, including adiabatic elimination and Born-Markov master equations, can then be derived as a particular case of the Nakajima-Zwanzig equation. Understanding the projector approach to open quantum systems is a powerful asset not only to derive effective dynamics in any open quantum system, but also to understand the deep connection between all the known methods and to help clarify confusing nomenclature conflicts across different fields~\footnote{Most prominently, the term ``adiabatic elimination'' refers to slightly different procedures in the atomic physics and the optomechanics communities~\cite{BrionJPhysA2007,WilsonRaeNJP2008}, a past source of confusion for the author and the spark that ignited this tutorial.}.

This tutorial aims at providing a practical guide for quantum theorists on
 how to rigorously derive reduced dynamics in common physical scenarios and regimes.  This tutorial summarizes well-known results in open quantum systems theory, and assumes the reader has some experience with open quantum systems and the derivation of basic master equations. We emphasize that, despite the apparent mathematical complexity, this tutorial aims only at providing a practical toolbox. It is thus not intended as a comprehensive review, nor as a complete bibliographical resource, nor as the most general analysis of the vast field of open quantum systems. For more general or more in-depth information we address the reader to other books~\cite{BreuerPetruccione,weiss2012quantum,CarmichaelBook,GardinerZollerQNoise} or guides on open quantum systems~\cite{tutorial1,tutorial3,tutorial4,tutorial5,tutorial6,tutorial7}.

 This tutorial is organized as follows. First, we summarise the derivation of the fundamental equation governing any open system, namely the Nakajima-Zwanzig equation, in Sec.~\ref{SecNakajima}. Then, in Sec.~\ref{secsimplification}, we derive useful simplifications of such equation in the common cases of weak system-bath coupling, time-independent Liouvillians, and coherent coupling between system and bath. We then show how to recover three different common situations, namely the Brownian master equation, the Born-Markov master equation, and the atomic physics version of adiabatic elimination, by perturbative expansion  of the Nakajima-Zwanzig equation, in Secs. \ref{SecBrownian}, \ref{SecBornMarkov}, and \ref{SecAMOAE} respectively. Each of these sections contains detailed examples of relevance to quantum science, specifically oscillator displacement noise, optomechanical sideband cooling master equation, Purcell effect in cavity QED, and the adiabatic elimination in a Lambda system. Section~\ref{SecConclusions} is devoted to the conclusions. Additionally, in Appendix ~\ref{appendix} we provide a deeper insight into more advanced concepts such as bath two-time correlators and the quantum regression formula.

 \section{Statement of the problem and the Nakajima-Zwanzig equation}\label{SecNakajima}

\begin{figure}[t]
	\centering
	\includegraphics[width=0.8\linewidth]{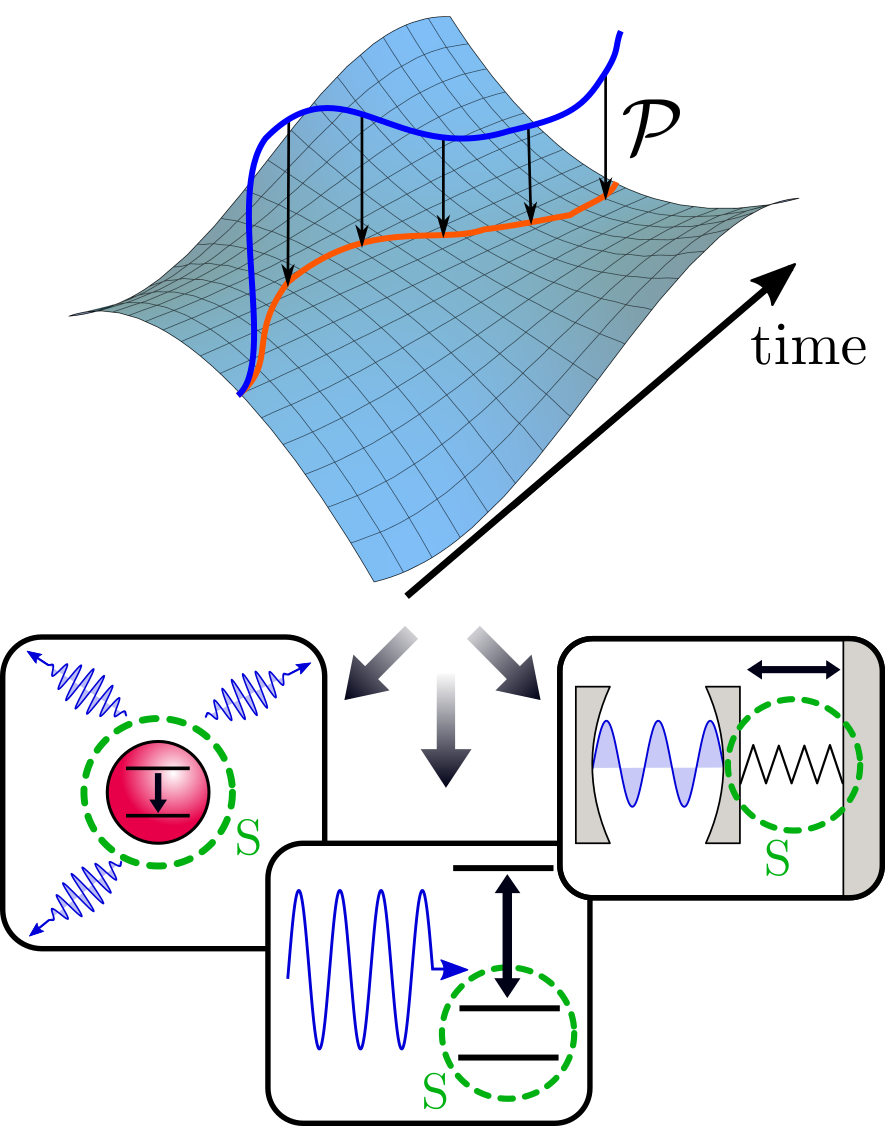}
	\caption{
	Schematic depiction of the projector approach: the time evolution of a system density matrix in the full Hilbert space (blue curve) is projected by a projector $\Py$ onto a subspace of interest (blue surface). As shown below, this subspace typically corresponds to product states between the reduced density matrix of a chosen set of degrees of freedom -- the ``system'' $S$ -- and a suitably chosen density matrix for the remaining ones -- the ``bath'' $B$. The dynamical equation within this subspace describes the evolution of the system (orange curve) influenced by the bath. This general and exact equation [the Nakajima-Zwanzig equation Eq.~\eqref{NZoriginal}] can be perturbatively expanded to recover various conventional methods to describe reduced dynamics, e.g., the Born-Markov equation for a system coupled to a continuum (left box, the continuum is represented by the blue arrows), the adiabatic elimination of levels in an $N-$level system (lower box), or the cooling of a mechanical resonator coupled to an optical cavity (right box).
	}
	\label{fig1}
\end{figure}

 We consider a quantum system with density matrix $\rho$, whose evolution is governed by a possibly time-dependent Liouvillian superoperator $\mathcal{L}(t)$~\cite{Haake1973,GardinerZollerQNoise}\footnote{In the first studies of stochastic quantum dynamics and dissipation the Liouvillian was originally encoded in the so-called dynamical matrices~\cite{SudarshanPR1961}},
\begin{equation}\label{firstEq}
    \dot{\rho} = \Lv(t) \rho.
\end{equation}
Our goal, as in any open quantum systems scenario, is to determine the effective dynamics of only a part of the whole system. Typically we call the part of interest ``system'' $S$ (or relevant part) and the remaining part ``bath'' $B$ (or irrelevant part)~\cite{BreuerPetruccione,Haake1973,FischerPRA2007}. The effective dynamics of the system under the action of the bath depends heavily on the case under consideration but, at the fundamental level, can be formulated in the same way for all cases. In this section we present this general formulation, which is the starting point to derive, in the following sections, the most common forms of effective dynamics found in quantum science.

In order to derive the effective dynamics of a part of the system, we start by 
defining a projector $\Py=\Py^2$ in the space of density matrices, and its complementary projector $\Qy=\Qy^2$:
\begin{equation}\label{Qdefinition}
    \Qy \equiv \Idy - \Py.
\end{equation}
Here, $\Idy$ represents the identity superoperator. Note that, throughout this tutorial, we will use normal font for operators acting on the space of kets/wavefunctions (e.g., $\hat{H}, \Id...$) and cursive font for the superoperators acting on the space of density matrices (e.g., $\Py$, $\Qy$, $\Idy$...).  
These two projectors are orthogonal, i.e.,
\begin{equation}
    \Py \Qy = \Qy \Py = 0,
\end{equation}
and can be used to define two contributions to the density matrix as follows~\cite{Haake1973},
\begin{equation}
    \rho = \Py \rho + \Qy \rho \equiv v + w.
\end{equation}
The motivation to split the Hilbert space in two parts using the projectors $\Py$ and $\Qy$ is intimately connected to the distinction between system and bath in open quantum systems.
As we will see in the next sections and schematically depict in Fig.~\ref{fig1}, we will typically (though not necessarily) identify $\Py$ with a projection onto product states between a suitably chosen bath density matrix and a density matrix of the system $\text{Tr}_B[v]$ (here $\text{Tr}_B$ indicates partial trace over the bath degrees of freedom) which, if the projectors are chosen properly and the system parameters are in the proper regime, will be an approximation to the true reduced density matrix $\text{Tr}_B[\rho]$~\cite{FischerPRA2007,GardinerZollerQNoise}. Therefore, at the general level considered in this section, our goal is to derive a dynamical equation for the contribution $v$ that depends only on $v$ itself. We remark that the projector $\Py$ must be chosen differently for each situation depending on the system.

To derive an equation of evolution for $v$, we first introduce the identity superoperator $\Py + \Qy$ both to the right and to the left of the Liouvillian in Eq. \eqref{firstEq}.  This allows us to split this equation into two orthogonal coupled equations,
\begin{equation}\label{evolutionV}
    \dot{v} = \Py\Lv v + \Py \Lv w,
\end{equation}
\begin{equation}\label{evolutionW}
    \dot{w} = \Qy\Lv v + \Qy \Lv w.
\end{equation}
Here we have assumed the projectors are time-independent so that $\mathcal{P}\dot{\rho} = \dot{v}$. Although time-independent projectors are usual in open quantum systems where system-bath correlations are small, we remark that a formalism employing time-dependent projection operators can be used to describe cases beyond this paradigm, e.g., in quantum optics~\cite{WillisPRA1974,PicardPRA1977}, Brownian motion~\cite{MoriPTP1965}
and non equilibrium statistical mechanics~\cite{RobertsonPR1966,KawasakiPLA1972,OchiaiPLA1973}.

Our second step is to formally solve Eq. \eqref{evolutionW} and introduce it in Eq. \eqref{evolutionV}. We do so by defining a propagator $\mathcal{G}(t,t_0)$ fulfilling~\cite{WilsonRaeNJP2008,BreuerPetruccione,Haake1973}
\begin{equation}\label{EquationG}
    \dot{\mathcal{G}}(t,t_0) = \Qy\Lv(t)\mathcal{G}(t,t_0) \hspace{0.4cm} ; \hspace{0.4cm} \mathcal{G}(t_0,t_0)=\Idy.
\end{equation}
Here $t_0$ is an arbitrary initial time.
From the identity $\mathcal{G}(t,t')\mathcal{G}^{-1}(t,t')=\Idy$ we can also derive the properties of its inverse,
\begin{equation}
    \dot{\mathcal{G}}^{-1}(t,t_0) = -\mathcal{G}^{-1}(t,t_0)\Qy\Lv(t),
\end{equation}
\begin{equation}
    \mathcal{G}^{-1}(t_0,t_0)=\Idy.
\end{equation}
Using these definitions one can cast Eq. \eqref{evolutionW} in the form
\begin{equation}
    \frac{d}{dt}(\mathcal{G}^{-1}(t,t_0)w(t)) = \mathcal{G}^{-1}(t,t_0)\Qy\Lv(t) v(t),
\end{equation}
and formally solve it as~\cite{Haake1973}
\begin{multline}\label{wformalsol}
    w(t) = \mathcal{G}(t,t_0)\Big[w(t_0)+
    \\
    +\int_{t_0}^tdt'\mathcal{G}^{-1}(t',t_0)\Qy\Lv(t')v(t')\Big].
\end{multline}
Hereafter we will assume
\begin{equation}\label{assumptionW}
   w(t_0) = 0.
\end{equation}
In open quantum systems, where one aims at describing the dynamics of a system coupled to a bath, one typically chooses projectors which map into product states, i.e., projectors such that $\mathcal{P}\rho(t) = \rho_B\otimes\rho_S(t)$, with $\rho_B$ a stationary density matrix of the bath and $\rho_S$ a density matrix on the system subspace~\cite{WillisPRA1974,BreuerPetruccione,GardinerZollerQNoise} (see Sec.~\ref{secsimplification} for details). For this kind of projectors, Eq.~\eqref{assumptionW} is equivalent to assuming no initial correlations between system and bath since, if $\rho(t_0) = \rho_B\otimes\rho_S(t_0)$, then $w(t_0) = (1-\Py)\rho(t_0) = 0$~\cite{BreuerPetruccione,toda2012statistical,Haake1973}. Analyzing the implications of initial system-bath correlations is beyond the scope of this tutorial, but we remark that this is a relevant and complex topic in the field of open quantum systems (see e.g.~\cite{Colla2022} and references therein). Introducing Eq.~\eqref{wformalsol} in Eq.~\eqref{evolutionV} we obtain the desired equation,
\begin{multline}\label{NakajimaZwanzigSymbolic}
    \dot{v}(t) = \Py\Lv(t) v(t)+
    \\
    \Py\Lv(t) \mathcal{G}(t,t_0)\int_{t_0}^tdt'\mathcal{G}^{-1}(t',t_0)\Qy\Lv(t')v(t').
\end{multline}
We can write a more explicit expression by explicitly solving for the propagator $\mathcal{G}(t,t_0)$. Formally integrating Eq.~\eqref{EquationG} yields
\begin{equation}
    \mathcal{G}(t,t_0) = \Idy +\int_{t_0}^t dt'\Qy \Lv(t')\mathcal{G}(t',t_0).
\end{equation}
By repeatedly reintroducing this equation onto itself we obtain a solution in terms of an infinite series~\cite{Haake1973},
\begin{widetext}
 \hspace{-3cm}\begin{equation}
\begin{split}
   \mathcal{G}(t,t_0) &= \Bigg(\Idy +\int_{t_0}^t dt' \Qy \Lv (t')  
     +\int_{t_0}^t dt'\int_{t_0}^{t'} dt'' \Qy \Lv (t') \Qy \Lv (t'')+...  \Bigg)=
    \\
    &
    =
    \Bigg(\Idy +\int_{t_0}^t dt' \mathcal{T}_+\left[\Qy \Lv (t')  \right]
    +\frac{1}{2!}\int_{t_0}^t dt'\int_{t_0}^{t} dt'' \mathcal{T}_+\left[\Qy \Lv (t') \Qy \Lv (t'')\right]+...  \Bigg)=
    \\&=
     \mathcal{T}_+\left[\exp\int_{t_0}^t dt' \Qy\Lv(t')\right]
    ,
\end{split}
\end{equation}
\end{widetext}
%\end{widetext}
where $\mathcal{T}_+$ is the usual time-ordering superoperator~\cite{Haake1973} (for a detailed definition and discussion in the context of quantum time-evolution see e.g. Ref.~\cite{fetter2012quantum}, Chapter 3). Similarly,
\begin{equation}
    \mathcal{G}^{-1}(t,t_0)=\mathcal{T}_-\left[\exp\left(-\int_{t_0}^t dt' \Qy\Lv(t')\right)\right],
\end{equation}
with $\mathcal{T}_-$ the time anti-ordering superoperator. 
Introducing the above two expressions into Eq.~\eqref{NakajimaZwanzigSymbolic}, using the composition identity for the time propagators~\cite{BreuerPetruccione}, $\mathcal{G}(t_3,t_1)=\mathcal{G}(t_3,t_2)\mathcal{G}(t_2,t_1)$ for $t_1<t_2<t_3$, and taking $t_0=0$ without loss of generality, we obtain the explicit closed equation for the density matrix of interest, $v(t)$,
% \begin{equation}\label{NZoriginal}
%     \boxed{
%     \dot{v}(t) =\Py\Lv(t) v(t) + \Py \Lv(t)\int_{0}^t d\tau \mathcal{T}_+\left[\exp\int_\tau^t dt' \Qy\Lv(t')\Qy\right]\Qy\Lv (\tau) v(\tau).
%     }
% \end{equation}
\begin{multline}\label{NZoriginal}
    \dot{v}(t) =\Py\Lv(t) v(t) + \Py \Lv(t)\int_{0}^t d\tau 
    \\
    \mathcal{T}_+\left[\exp\int_\tau^t dt' \Qy\Lv(t')\Qy\right]\Qy\Lv (\tau) v(\tau).
\end{multline}
In the above equation we have added trivial $\Qy$ projectors to the exponentials to obtain a symmetric expression. This equation is known as the Nakajima-Zwanzig equation~\cite{NakajimaOriginal,ZwanzigOriginal,MoriPTP1965}, and is completely general aside from the two assumptions, namely $w(0)=0$ and time-independent projector $\mathcal{P}$. Note that the Nakajima-Zwanzig equation is trivially fulfilled for the two limiting cases of $\Py = \Idy$ (i.e., the subspace of interest is the whole space of density matrices) and $\Py = 0$ (i.e., the complementary subspace is is the whole space of density matrices). 
We remark that one of the most complex aspects in solving Eq.~\eqref{NZoriginal} is its ``memory'', that is, the dependence of $v(t)$ on its values at all previous times, captured by the term $v(\tau)$ (with $0<\tau<t$) within the integral. Although such time non-locality, also known as non-Markovianity, is relevant in several specific systems such as sub-Ohmic spin-boson models~\cite{GonzalezBallesteroPRB2017}, Brownian motion~\cite{GroblacherNatComm2015}, and others (see~\cite{deVegaRevModPhys2017} and refs. therein), it remains beyond the scope of this tutorial. As discussed below, here we will focus on the common regimes of open quantum systems where a time-local approximation (Markov approximation) holds. As a final note of interest, note that the non-Markovianity of Eq.~\eqref{NZoriginal} can be ``hidden'' into the integral Kernel to obtain a non-Markovian, but time-local, equation for $v(t)$, usually named the time-convolutionless master equation~\cite{ShibataJSP1977,FULINSKI1968575,BreuerPetruccione}.

It is worth noting that the solution of Eq.~\eqref{NZoriginal} involves the full solution of the whole system, that is, the Nakajima-Zwanzig equation is just a recasting of Eq. \eqref{firstEq} and has the same complexity. This recasting is particularly convenient as it allows for a more controlled and systematic perturbative treatment and, in specific cases, for an exact solution. For this reason the Nakajima-Zwanzig equation has been employed to describe multiple complex problems, such as central spin models~\cite{FischerPRA2007,FerraroPRB2008,BarnesPRL2012} (both beyond-weak coupling and exact solutions), element synthesis processes in heavy ion collisions~\cite{AbePTEP2021}, stochastic dynamics of nonlinear systems~\cite{VenturiPRSA2014}, and charge transport~\cite{YangJCP2019}, among others. In the context of open quantum systems, the focus of this tutorial, the true strength of the Nakajima-Zwanzig equation relies on its potential for a perturbative expansion of the Liouvillian in terms of the relevant system and bath timescales, as we will see below.

\section{Simplification of Nakajima-Zwanzig equation for usual open quantum systems}\label{secsimplification}

In this section we formulate the Nakajima-Zwanzig Eq.~\eqref{NZoriginal} in terms of system, bath, and interaction Liouvillians, and derive simplified expressions for  such equation under common assumptions and regimes in open quantum systems.

\subsection{Usual form of the Liouvillian}

In the context of open quantum systems, the total Hilbert space $\mathbb{H}$ is usually split as $\mathbb{H}=\mathbb{H}_S\otimes \mathbb{H}_B$ where $\mathbb{H}_S$ and $\mathbb{H}_B$ describe the individual Hilbert spaces for the system and the bath respectively. The Liouvillian can thus also be written as a sum of three terms~\cite{BreuerPetruccione,Haake1973},
\begin{equation}\label{generalLiouv}
    \Lv(t) = \Lv_S(t) + \Lv_B(t) + \Lv_{\text{Int}}(t),
\end{equation}
corresponding to the Liouvillians of the system degrees of freedom (whose reduced dynamics we aim at deriving), the bath degrees of freedom, and their interaction respectively. We choose to define $\Lv_S(t)$ and $\Lv_B(t)$ as the terms in the Liouvillian that act only on the subspaces of system and bath, respectively. That is, if one writes a general operator $\{\hat{O}:\mathbb{H}\to\mathbb{H}\}$ as $\hat{O}=\sum_\alpha \hat{S}_\alpha\otimes\hat{B}_\alpha$ with $\hat{S}_\alpha$ and $\hat{B}_\alpha $ arbitrary system and bath operators (i.e., $\hat{S}_\alpha:\mathbb{H}_S\to \mathbb{H}_S$ and $\hat{B}_\alpha:\mathbb{H}_B\to \mathbb{H}_B$) and $\alpha$ an arbitrary index, then by definition
\begin{equation}\label{LSdiagonal}
    \Lv_S (t) \sum_\alpha \hat{S}_\alpha\otimes\hat{B}_\alpha = \sum_\alpha \left(\Lv_S(t) \hat{S}_\alpha\right)\otimes\hat{B}_\alpha,
\end{equation} 
and 
\begin{equation}\label{LBdiagonal}
    \Lv_B(t)\sum_\alpha \hat{S}_\alpha\otimes\hat{B}_\alpha=\sum_\alpha \hat{S}_\alpha\otimes\left(\Lv_B(t) \hat{B}_\alpha\right).
\end{equation}
Conversely, all the terms acting non-trivially on both $S$ and $B$ are by definition contained in $\Lv_{\text{Int}}(t)$. This term in the Liouvillian thus describes the coupling between the system and the bath degrees of freedom.

Usually, the projector $\Py$ is chosen such that it projects onto product state density matrices with a common bath state, i.e,~\cite{Haake1973,BreuerPetruccione,GardinerZollerQNoise,toda2012statistical}
\begin{equation} \label{Pydefinition}
    \Py :\to \Py (*) =  \text{Tr}_B\left[(*)\right] \otimes \rho_B,
\end{equation}
where $(*)$ indicates the argument operator and $\text{Tr}_B$ a partial trace over the bath degrees of freedom. Note that in order for the above operator to be a projector ($\mathcal{P}=\mathcal{P}^2$) the state $\rho_B$ must be a physically consistent quantum state fulfilling $\text{Tr}_B[\rho_B]=1$.
By definition
\begin{equation}\label{vprov}
    v(t) = \mathcal{P}\rho = \rho_S(t)\otimes\rho_B,
\end{equation}
where $\rho_S(t)$ is 
the density matrix of the system by definition. As we will see below, under the weak system-bath coupling approximation and for an appropriately chosen projector $\Py$, Eq.~\eqref{vprov} approximates the total density matrix, i.e. $\rho(t) \approx v(t)$~\cite{FischerPRA2007}. The choice of $\rho_B$ in Eq.~\eqref{Pydefinition} is physically motivated by the problem at hand. The most popular choice in open quantum systems is a steady state of the bath Liouvillian~\cite{BreuerPetruccione,GardinerZollerQNoise,FischerPRA2007}, 
\begin{equation}\label{steadystatechoice}
    \Lv_B(t)\rho_B = 0.
\end{equation}
This choice makes $v(t)$ a good approximation for the total density matrix if the bath rapidly decays to its steady state or if the bath is very large (i.e., it contains a large number of modes as compared with the system), two common situations in conventional open quantum systems~\cite{BreuerPetruccione,GardinerZollerQNoise,FischerPRA2007}. Throughout this tutorial we will assume the choice Eq.~\eqref{steadystatechoice}. We remark, however, that more complicated situations in open quantum systems require the choice of more involved projectors and/or different (and possibly time-dependend) bath states $\rho_B$, for instance if part of the system-induced bath evolution~\cite{IgnatyukArxiv2019} or non-negligible system-bath correlations~\cite{FischerPRA2007,FerraroPRB2008,BarnesPRL2012} must be taken into account.

For the projector defined in Eq. \eqref{Pydefinition}, the following general properties can be demonstrated~\cite{Haake1973,GardinerZollerQNoise}:
\begin{equation}\label{PLscomm}
    \Lv_S(t)\Py=\Py \Lv_S(t) ,
\end{equation}
\begin{equation}\label{PLBeq0}
    \Lv_B(t)\Py =0  = \Py \Lv_B(t).
\end{equation}
The first of these properties follows from Eq. \eqref{LSdiagonal}. Regarding the second property, 
 the first equality is a consequence of Eq. \eqref{steadystatechoice},
whereas the second equality can be derived assuming $\mathcal{L}_B$ is a physical (i.e., trace- and positivity-preserving) Liouvillian~\footnote{ A Liouvillian preserving trace and positivity is the infinitesimal generator of a completely positive trace-preserving dynamical map~\cite{BreuerPetruccione}, and can in general be written as~\cite{bernal2019multiple}
\begin{multline}
    \mathcal{L}_B[(*)] = -i[\hat{h}(t),(*)] 
    \\+ \sum_\alpha \gamma_\alpha(t)\left[\hat{B}^\dagger_\alpha (*) \hat{B}_\alpha - \frac{1}{2}\{\hat{B}_\alpha\hat{B}^\dagger_\alpha,(*)\}\right].
\end{multline}
By using this expression it is straightforward to prove that $\Py\mathcal{L}_B=0$
.}. In many cases, as we will see in the following sections, it is also possible to demonstrate that $\Py\Lv_{\rm Int}(t)\Py=0$~\cite{Haake1973,GardinerZollerQNoise,BreuerPetruccione}, but since this is not always the case we will not assume this equality.

The above properties allow for a simplification of the Nakajima-Zwanzig equation~\cite{BreuerPetruccione}. After taking the partial trace over the bath we find
\begin{multline}\label{Nakajimasimplified}
    \dot{\rho}_S(t) =\Lv_S(t)\rho_S(t) +\Py\Lv_{\rm Int}(t)\Py+
    \\
    \text{Tr}_B  \Lv_{\rm Int}(t)\int_0^t d\tau \mathcal{T}_+\left[\exp\int_\tau^t dt' \Qy\Lv(t')\Qy\right]
    \\
    \Qy\Lv_{\rm Int}(\tau) v(\tau).
\end{multline}
%\begin{widetext}
%\begin{equation}\label{Nakajimasimplified}
%    \dot{\rho}_S(t) =\Lv_S(t)\rho_S(t) + \text{Tr}_B\left[\Py\Lv_{\rm Int}(t) v(t)\right] + 
%    \text{Tr}_B  \Lv_{\rm Int}(t)\int_0^t d\tau \mathcal{T}_+\left[\exp\int_\tau^t dt' \Qy\Lv(t')\Qy\right]\Qy\Lv_{\rm Int}(\tau) v(\tau).
%\end{equation}
%\end{widetext}
% \begin{widetext}
% \begin{equation}\label{Nakajimasimplified}
% \begin{split}
%     \dot{\rho}_S(t) &=\Lv_S(t)\rho_S(t) + \text{Tr}_B\left[\Py\Lv_{\rm Int}(t) v(t)\right] + 
%     \\
%     &
%     +\text{Tr}_B  \Lv_{\rm Int}(t)\int_0^t d\tau \mathcal{T}_+\left[\exp\int_\tau^t dt' \Qy\Lv(t')\Qy\right]\Qy\Lv_{\rm Int}(\tau) v(\tau).
% \end{split}
% \end{equation}
% \end{widetext}

\subsection{Weak coupling and time-independent system \& bath Liouvillians}\label{subsecWC}

Equation~\eqref{Nakajimasimplified} is still in general difficult to solve, but can be expanded perturbatively.
There are different methods for this perturbative expansion (some explained in detail in the next sections) but  in the context of open quantum systems they all share the assumption of \textit{weak system-bath coupling}~\cite{toda2012statistical}. 
Indeed, the paradigm of open quantum systems, namely a well-defined system whose free dynamics are modified in a relatively small way (e.g., by the addition of damping) due to coupling to a well-defined bath, only makes sense if the coupling is weak. Otherwise system and bath hybridize strongly and both their dynamics are completely different from their uncoupled analogues. We remark that this is not always the case~\cite{LednevPRL2024}, and address the reader to Refs.~\cite{BeaudoinPRA2011,ChiangPRA2021,BambaPRA2014,SettineriPRA2018,BeckerPRL2022} for more information about the theoretical approach to strong system-bath coupling in specific quantum platforms.
Until we introduce more rigorous definitions in the next sections, we can say system and bath are weakly coupled when the timescales introduced by the interaction Liouvillian $\mathcal{L}_{\rm Int}$ (typically coupling rates) are much longer (typically $\epsilon^{-1}$ times longer, with $\epsilon \ll 1$ a dimensionless coupling strength parameter) than the timescales associated to $\mathcal{L}_S + \mathcal{L}_B$. In this situation we can neglect the interaction Liouvillian within the exponential in Eq.~\eqref{Nakajimasimplified}~\cite{toda2012statistical,GardinerZollerQNoise}, 
i.e.,
\begin{widetext}
\begin{multline}
    \mathcal{T}_+\left[\exp\int_\tau^t dt' \Qy\Lv(t')\Qy\right] = \mathcal{T}_+\left[\exp\int_\tau^t dt' \Qy(\mathcal{L}_B(t')+\Lv_S(t'))\Qy\right]\left(\Idy + \mathcal{O}\left(\epsilon\right)\right)
    \\\approx \mathcal{T}_+\left[\exp\int_\tau^t dt' \Qy(\mathcal{L}_B(t')+\Lv_S(t'))\Qy\right].
\end{multline}
This allows us to write our first approximated Nakajima-Zwanzig equation as~\cite{GardinerZollerQNoise,BreuerPetruccione}
\begin{multline}\label{NakajimasimplifiedWC}
    \dot{\rho}_S(t) \approx\Lv_S(t)\rho_S(t) + \text{Tr}_B\left[\Py\Lv_{\rm Int}(t) v(t)\right] + \\+
    \text{Tr}_B  \Lv_{\rm Int}(t)\int_0^t d\tau \mathcal{T}_+\left[\exp\int_\tau^t dt' \left(\Lv_B(t')+\Lv_S(t')\right)\right]\Qy\Lv_{\rm Int}(\tau) v(\tau).
\end{multline}
\end{widetext}
Here we have used the fact that (i) due to the projector $\Qy$ appearing after the exponential, we can substitute $\Qy (\mathcal{L}_B+\Lv_S)\Qy \to \Qy (\mathcal{L}_B+\Lv_S)$, (ii) the projector $\Qy$ can be shown, from Eqs.~\eqref{PLscomm} and~\eqref{PLBeq0}, to commute with both Liouvillians, i.e., $\Qy (\mathcal{L}_B+\Lv_S) = (\mathcal{L}_B+\Lv_S)\Qy$, and (iii) again due to the projector $\Qy$ after the exponential we can substitute   $(\mathcal{L}_B+\Lv_S)\Qy  \to \mathcal{L}_B+\Lv_S$. 

 A second simplification can be undertaken by noting that in many problems of interest the system and bath Liouvillian $\Lv_S$ and $\Lv_B$ are time-independent, or can be made time-independent by a suitable transformation (e.g., by transforming to the interaction picture). In this case the above expression simplifies to
 
  \vspace{0.3cm}
 \noindent\fbox{\begin{minipage}{\dimexpr\linewidth-2\fboxsep-2\fboxrule\relax}
 \centering
 \vspace{-0.3cm}
 \begin{multline}\label{NakajimasimplifiedTime}
     \dot{\rho}_S(t) \approx\Lv_S\rho_S(t) + \text{Tr}_B\left[\Py\Lv_{\rm Int}(t) v(t)\right] + \\+ 
     \text{Tr}_B  \Lv_{\rm Int}(t)
     \int_0^t d\tau 
 e^{\left[\Lv_B+\Lv_S\right](t-\tau)}\Qy\Lv_{\rm Int}(\tau) v(\tau).
 \end{multline}
 \end{minipage}}

  \vspace{0.3cm}
\noindent 
Equation~\eqref{NakajimasimplifiedTime} will be the basis for the rest of the tutorial. Specifically, in the following sections, we will address how to derive effective system dynamics from this equations in three situations of particular interest in applied quantum theory.

\subsection{Coherent interaction}\label{SecCoherentInteraction}

Although not the most general case, common introductions to open quantum systems as well as the examples in this tutorial focus on a system coupled to a bath via a weak and fully coherent (i.e., purely Hamiltonian) system-bath interaction. This is the case for e.g. conventional derivations of the Born-Markov master equation. In this section we write explicitly the Nakajima-Zwanzig equation for this situation in order to make a connection with more conventional approaches. We emphasize that this section has an illustrative purpose but is
not essential to follow the rest of the tutorial. Note also that the derivations in this tutorial are general and apply as well to dissipative system-bath interactions, which typically arise when system and bath are coupled indirectly via their common interaction with a third reservoir (see e.g. Refs.~\cite{Mogilevtsev2015,GonzalezTudelaPRL2011} for examples of dissipative coupling in quantum and atom optics).

We consider a system and a bath with Liouvillians given by
\begin{equation}
    \Lv_{S}^{(0)}(t) (*) = -\frac{i}{\hbar}\left[\hat{H}_S(t),(*)\right] + \mathcal{D}_S(t)[(*)],
\end{equation}
\begin{equation}
    \Lv_B (*) = -\frac{i}{\hbar}\left[\hat{H}_B,(*)\right]+\mathcal{D}_B[(*)],
\end{equation}
where we use the super-index ``$(0)$'' to indicate the original Liouvillians before we apply any redefinition (see below). In the expressions above $\hat{H}_S(t)$ and $\hat{H}_B$ are the system and bath Hamiltonians and
$\mathcal{D}_S(t)[(*)]$ and $\mathcal{D}_B[(*)]$ are arbitrary dissipators, i.e., the contributions to system and bath evolutions that cannot be described as a coherent Hamiltonian evolution (e.g., system or bath decay into external reservoirs).
The interaction Liouvillian is assumed to be purely conservative, i.e., it includes only a coherent system-bath interaction given by the interaction Hamiltonian $\hat{V}^{(0)}$,
\begin{multline}\label{LintMEQ}
    \Lv_{\rm Int}^{(0)}(t) (*) =
    -\frac{i}{\hbar}\left[\hat{V}^{(0)},(*)\right]
    \\=-\frac{i}{\hbar}\left[\hbar\sum_\alpha\hat{S}_\alpha(t)\otimes\hat{B}^{(0)}_{\alpha}(t),(*)\right],
\end{multline}
with $\hat{S}_\alpha(t)=\hat{S}^\dagger_\alpha(t)$ and $\hat{B}^{(0)}_{\alpha}(t)=\hat{B}^{(0)\dagger}_{\alpha}(t)$ arbitrary system and bath operators.

\begin{table*}[t!]
\centering
 \begin{tabularx}{\linewidth}{ 
  | >{\raggedright\arraybackslash}X | >{\raggedright\arraybackslash}X 
  | >{\raggedright\arraybackslash}X 
  | >{\raggedright\arraybackslash}X | } 
 Particular case  & Section~\ref{SecBrownian} & Section~\ref{SecBornMarkov} & Section~\ref{SecAMOAE}  \tabularnewline [1.5ex] 
 \hline\hline
 Characteristics & Bath decays faster than any other timescale & Bath decays faster than any other timescale except system oscillation frequencies & Bath oscillates faster than any other timescale  \tabularnewline  [1ex] 
 \hline
  Validity regime & 
  $\min(\text{Re}[\lambda_j^B]) \gg\{\text{Im}[\lambda_j^B],\vert\lambda_j^S\vert,\vert\lambda_j^{\rm Int}\vert \}$
  & 
  $\min(\text{Re}[\lambda_j^B])
      \gg\{\text{Im}[\lambda_j^B],\text{Re}[\lambda_j^S],\vert\lambda_j^{\rm Int}\vert \}$
 & 
 $\min(\text{Im}[\lambda_j^B])
 \gg\{\text{Re}[\lambda_j^B],\vert\lambda_j^S\vert,\vert\lambda_j^{\rm Int}\vert \}$
 \tabularnewline  [1ex] 
 \hline
 Usual nomenclature & Brownian master equation & Born-Markov master equation, adiabatic elimination (in optomechanics \& sideband cooling contexts) & Adiabatic elimination (in multilevel atom optics), [truncated] Schrieffer-Wolff transformation  \tabularnewline  [1ex] 
 \hline
 \end{tabularx}
 \caption{Summary of the three particular expansions of the Nakajima-Zwanzig equation that we consider in this tutorial. In the third row the quantities $\lambda_j^X$ label the eigenvalues of the different terms of the Liouvillian, $\mathcal{L}_X$, see Sec.~\ref{SecBrownian} for details and definition of Liouvillian eigenvalues.}\label{Tableparams}
\end{table*}

In the most typical approach, namely the Born-Markov master equation (see Sec.~\ref{SecBornMarkov}), it is usual to assume that the bath degrees of freedom decay faster than the system-bath interaction rate, but possibly slower than the system and bath free oscillation frequencies~\cite{weiss2012quantum,GardinerZollerQNoise,BreuerPetruccione}. In such case it is usual to transform to the interaction picture in order to keep these frequencies implicit and avoid neglecting terms proportional to them in the subsequent approximations. However, the transformation to the interaction picture should be preceded by the crucial step of ensuring that the bath operators have zero expectation value over the bath steady state, i.e., $\langle \hat{B}_\alpha^{(0)}(t)\rangle_{\rm ss} \equiv \text{Tr}_B[\rho_B\hat{B}^{(0)}_\alpha(t) ] = 0$~\cite{GardinerZollerQNoise}. If this is not the case, the integral in Eq.~\eqref{NakajimasimplifiedWC} might diverge (this occurs, for instance, in the case of closed system and bath, $\mathcal{D}_S^{(i)}=\mathcal{D}_B^{(i)}=0$). 
This divergence stems from the fact that the perturbative expansion includes a part of the interaction $\hat{V}^{(0)}$ which is purely a system operator, namely the driving term $\text{Tr}_B[\rho_B \hat{V}^{(0)}(t)] = \hbar \sum_\alpha \hat{S}_\alpha(t)\langle\hat{B}^{(0)}_\alpha(t)\rangle_{\rm ss}$~\footnote{At a deeper level, in the language of Appendix~\ref{appendix}, this part of the interaction results in master equation terms that are not governed by the steady-state correlators of the bath $\langle \hat{B}_\alpha(t+\tau)\hat{B}_{\alpha'}(t)\rangle_{\rm ss}$, which typically decay as a function of $\tau$ and thus ensure the convergence of the integral. Instead, these terms depend on
the single expectation values $\langle \hat{B}_\alpha(t+\tau)\rangle\langle\hat{B}_{\alpha'}(t)\rangle_{\rm ss} = \langle \hat{B}_\alpha\rangle\langle\hat{B}_{\alpha'}\rangle_{\rm ss}$, which by definition are time-independent in the steady state (provided that $\Lv_B$ is time-independent) and thus do not provide the decaying behaviour needed for the integral to converge.}. 
To avoid this divergence, we need to redefine the system and interaction Liouvillians in the Schr\"odinger picture in the following way,
\begin{multline}
    \Lv_S(t) (*) = -\frac{i}{\hbar}\left[\hat{H}_S(t)+\hat{S}_V(t),(*)\right] +
    \\+\mathcal{D}_S(t)[(*)],
\end{multline}
\begin{multline}\label{Lintmodified}
    \Lv_{\rm Int}(t)(*) = -\frac{i}{\hbar}\left[\hat{V}^{(0)}(t)-\hat{S}_V(t),(*)\right]\\\equiv-\frac{i}{\hbar}\left[\hat{V}(t),(*)\right],
\end{multline}
where we have defined the system operator
\begin{equation}
    \hat{S}_V(t) \equiv \text{Tr}_B[\rho_B\hat{V}(t)],
\end{equation}
and the modified interaction potential
\begin{multline}\label{V0000}
    \hat{V}(t) \equiv \hat{V}^{(0)}(t)-\hat{S}_V(t)=
    \\
    =\hbar\sum_\alpha\hat{S}_\alpha\otimes\left[\hat{B}_\alpha^{(0)}-\langle \hat{B}_\alpha^{(0)}(t)\rangle_{\rm ss}\right],
\end{multline}
whose bath operators have zero expectation value by definition.
Note that, since the contributions added to $\Lv_S^{(0)}$ and $\Lv_{\rm Int}^{(0)}$ are equal in magnitude and have opposite sign, the global Liouvillian remains unchanged, i.e., $\Lv_S(t)+\Lv_{\rm Int}(t) = \Lv_S^{(0)}(t)+\Lv_{\rm Int}^{(0)}(t)$. However, by rearranging in this way  the interaction and the system Liouvillian, we can perform the perturbative expansion on system-bath couplings while avoiding divergences. Note also that after these redefinitions the interaction Liouvillian fulfills by definition the identity $\Py \Lv_{\rm Int}(t) \Py=0$.

After ensuring that the bath operators have zero expectation value, we can transform to the interaction picture with respect to $\hat{H}_S(t)+\hat{S}_V(t)+\hat{H}_B$, where the Liouvillians simplify to
\begin{equation}
    \Lv_j^{(i)} (*) = \mathcal{D}_j^{(i)} (t)[(*)] \hspace{0.3cm} (j=S,B),
\end{equation}
and the index $(i)$ denotes the interaction picture. For simplicity we will assume that the bath Liouvillian in the interaction picture is time-independent~\footnote{This is the case in general scenarios where system and bath Hamiltonians are diagonal and their dissipation corresponds to thermal decay/absorption, dephasing, and other common forms.}. Although this assumption is fulfilled in most usual scenarios (e.g., the quantum optical master equation), the following derivations can also be extended to time-dependent bath Liouvillians.

Under the above assumptions we can directly use Eq. \eqref{NakajimasimplifiedWC} to describe this situation,
\begin{multline}\label{NZMarkov}
    \dot{\rho}_S^{(i)}(t) \approx\mathcal{D}_S^{(i)}(t)[\rho_S(t)] + \text{Tr}_B  \Lv_{\rm Int}^{(i)}(t)\\
    \int_0^t d\tau e^{\mathcal{D}_B^{(i)}\tau} \mathcal{S}(t,\tau)
\Lv_{\rm Int}^{(i)}(t-\tau) v(t-\tau).
\end{multline}
Here we have changed integration variable from $\tau$ to $t-\tau$, made use of the fact that system and bath Liouvillians commute by definition, and defined the system evolution superoperator in the interaction picture
\begin{equation}
   \mathcal{S}(t,\tau) \equiv \mathcal{T}_+\left[\exp\int_0^\tau dt'\mathcal{D}_S^{(i)}(t-t')\right].
\end{equation}
To recover the common ``textbook'' expression of the master equation we introduce above the expression of the interaction Liouvillian Eq.~\eqref{Lintmodified} to obtain
\begin{multline}\label{NZMarkov22}
    \dot{\rho}_S^{(i)}(t) \approx\mathcal{D}_S^{(i)}(t)[\rho_S(t)] -\frac{1}{\hbar^2} \text{Tr}_B  \Big[\hat{V}^{(i)}(t),\\
    \int_0^t d\tau e^{\mathcal{D}_B^{(i)}\tau} \mathcal{S}(t,\tau)\left[
\hat{V}^{(i)}(t-\tau), v(t-\tau)\right]\Big].
\end{multline}
Equation \eqref{NZMarkov22} is
the usual expression for the Born Master equation for a dissipative system coupled to a dissipative bath~\cite{NavarreteQOCourse}.

In the next sections we go back to the more general representation~\eqref{NakajimasimplifiedTime} and consider three particular cases that are frequent in quantum optics and optomechanics, namely the Brownian master equation in Sec.~\ref{SecBrownian}, the Born-Markov master equation in Sec.~\ref{SecBornMarkov}, and the atom quantum optics version of adiabatic elimination in Sec.~\ref{SecAMOAE}. Their validity regimes and comparison to each other is shown in Table~\ref{Tableparams}.

\section{Particular case 1: Bath decays faster than any other timescale (Brownian master equation)}\label{SecBrownian}

Our first particular case consists of a system-bath interaction where the bath degrees of freedom decay much faster than any other timescale in the system. This situation is sometimes referred to as the Brownian Master Equation~\cite{BreuerPetruccione,GardinerZollerQNoise}. The most general Liouvillian corresponding to this case reads in the Schr\"odinger picture 
\begin{equation}\label{LiouvillianAE1}
    \Lv (t) = \xi^2\Lv_B +\xi\Lv_{\rm Int}(t) + \Lv_S(t),
\end{equation}
in terms of a large expansion parameter $\xi \gg 1$. The Liouvillian $\Lv_B$ is assumed time-independent (although not necessarily) and dominated by a dissipative contribution such that $\min (\text{Re}[\text{eigenvalues}[\xi^2\Lv_B]]) \propto \xi^2$~\cite{KesslerPRA2012,LiouvEinvals,Liouvsteadystate}. 
The dependences of $\mathcal{L}_{\rm Int}(t)$ and $\mathcal{L}_{S}(t)$ on $\xi$ do not need to be linear, but should grow slower than $\xi^2$ for the following expansion to be valid. Finally, note that any time dependence of $\mathcal{L}_{\rm Int}(t)$ and $\mathcal{L}_{S}(t)$ is also assumed much slower than the decay of $\Lv_B$.

Since by assumption $\Lv_B$ is time-independent and $\Lv_S$ is much smaller than $\Lv_B$, we can use directly Eq.~\eqref{NakajimasimplifiedTime} in the simplified form
\begin{multline}\label{NZreducedAM}
    \dot{\rho}_S(t) =\Lv_S(t)\rho_S(t)+\text{Tr}_B\left[\Lv_{\rm Int}(t) v(t)\right]
    \\
    + \text{Tr}_B \Lv_{\rm Int}(t)\int_0^t d\tau e^{\Lv_B \tau}\Qy \Lv_{\rm Int}(t-\tau) v(t-\tau) ,
\end{multline}
where we have changed integration variable from $t$ to $t-\tau$. 
We assume that a steady state of $\Lv_B$ exists, which implies the real part of its eigenvalues is negative (except for the eigenvalue $0$ corresponding to the steady state). This can be easily proven by vectorization of the bath von-Neumann equation for finite-size baths~\cite{vectorization1,vectorization2} and is generally true for common infinite baths appearing in open quantum systems~\cite{Liouvillianspectrum}. For a more detailed discussion on this issue and potential pathological cases (multiple steady states, no steady state, etc), see e.g. Refs.~\cite{LiouvEinvals,Liouvsteadystate} and references therein. Since the real part of the eigenvalues is negative, the function $\exp(\mathcal{L}_B\tau)$ decays as a function of $\tau$. To see this, one can formally define
right and left eigenvectors for the bath Liouvillian as follows~\cite{KesslerPRA2012}
\begin{equation}\label{righteigenvecs}
    \Lv_B \vert \mathbf{r}_j ) = \lambda_j \vert \mathbf{r}_j ) ,
\end{equation}
\begin{equation}\label{lefteigenvecs}
    ( \mathbf{l}_j \vert \Lv_B=  ( \mathbf{l}_j \vert \lambda_j,
\end{equation}
where $\lambda_j$ are the eigenvalues and where we use the different bracket notation $\vert )$ to distinguish Liouvillian eigenvectors (i.e. operators) from conventional kets. Note that by definition we exclude from this treatment any possible exceptional points of the bath spectrum, since at these points the Liouvillian is not diagonalizable~\cite{nonHermitian}.
These eigenvectors can be chosen biorthonormal $(\mathbf{l}_i\vert\mathbf{r}_k)=\delta_{ik}$ (with the overlap defined for their matrix representation as $(\mathbf{l}_i\vert\mathbf{r}_k)\equiv \text{Tr}[\mathbf{l}_i^\dagger \mathbf{r}_k]$) and forming a complete basis~\cite{KesslerPRA2012,LiouvEinvals}
\begin{equation}
    \sum_{j}\vert \mathbf{r}_j)(\mathbf{l}_j\vert = \mathbb{1}.
\end{equation}
The exponential of the bath Liouvillian acts on an arbitrary vector $\vert \varrho (t) )$ belonging to the subspace span by the above basis as
\begin{equation}\label{EqLvoverarbstate}
    e^{\xi^2\mathcal{L}_B\tau}\vert \varrho(t) ) =\sum_je^{\lambda_j \tau}( \textbf{l}_j\vert \varrho(t)) \hspace{1mm}\vert \textbf{r}_j).
\end{equation}
Note that the vector $\vert \varrho (t) )$ is not necessarily restricted to physical density matrices. Equation~\eqref{EqLvoverarbstate} confirms that, since $\text{Re}[\lambda_j]\le0$, $\exp(\mathcal{L}_B\tau)$ is a decaying function of $\tau$~\footnote{Note that we technically exclude the eigenvalue $\lambda_j=0$ in making this argument since, in usual scenarios -- and certainly in all examples of this tutorial -- it does not appear when evaluating the second line of Eq.~\eqref{NZreducedAM}. This is the case if the bath is formed by independent modes and $\rho_B$ represents a thermal state, as can be easily checked.}.
Moreover, it decays much faster than any other timescale, as by assumption $\text{Re}[\lambda_j]\propto \xi^2$ are the largest energy scales of the problem. 
Since both the density matrix $v$ and the interaction Liouvillian $\Lv_{\rm Int}$ evolve on a slower timescale ($\sim\xi$ at most), they practically remain unchanged during this decay. This justifies undertaking a Markov approximation, i.e., to remove the $\tau$ dependence in $v(t-\tau)$ and $\Lv_{\rm Int}(t-\tau)$ (the latter is the key difference with respect to the Born-Markov master equation in Sec.~\ref{SecBornMarkov}), and to extend the upper limit of the integral to infinity. We obtain
\begin{multline}\label{NZreducedBM}
    \dot{\rho}_S(t) \approx\Lv_S(t)\rho_S(t)+\text{Tr}_B\left[\Lv_{\rm Int}(t) v(t)\right]
    \\
    +\text{Tr}_B \Lv_{\rm Int}(t)\int_0^\infty d\tau e^{\Lv_B \tau}\Qy \Lv_{\rm Int}(t) v(t)
    \\
    = \Lv_S(t)\rho_S(t)+\text{Tr}_B\left[\Lv_{\rm Int}(t) v(t)\right]
    \\
    +\text{Tr}_B \Lv_{\rm Int}(t) \Lv_B^{(-1)}\Qy \Lv_{\rm Int}(t) v(t).
\end{multline}
This is the usual Brownian Master Equation found in the literature \cite{GardinerZollerQNoise,KesslerPRA2012}. We remark that, as we show in the examples in this tutorial, computing the eigenvalues of the bath Liouvillian is usually not needed to use the resulting master equations, since from the expressions of the Liouvillians it is typically straightforward to confirm that the hierarchy Eq.~\eqref{LiouvillianAE1} is fulfilled.

\subsection{Example: oscillator displacement noise}\label{brownianexample}

\begin{figure}[t]
	\centering
	\includegraphics[width=0.8\linewidth]{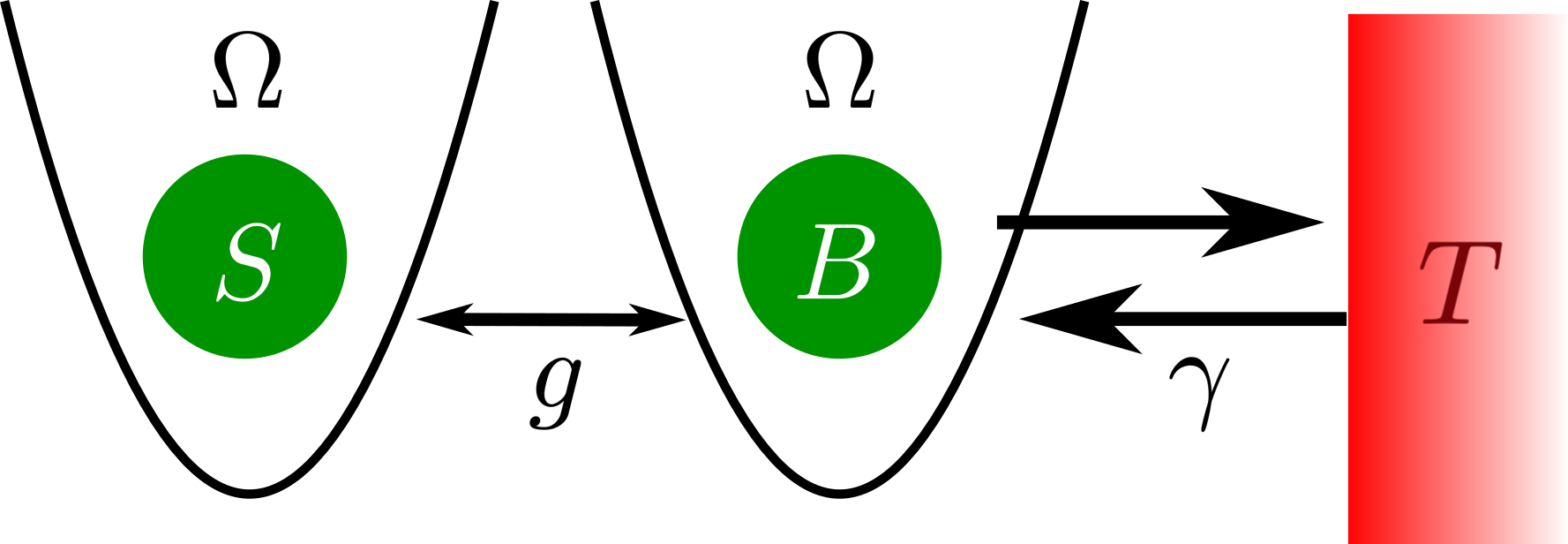}
	\caption{
	A ``system'' oscillator is coupled to a ``bath'' oscillator which decays to a thermal bath at a rate much faster than the oscillators' frequencies and coupling rates, $\gamma\gg\Omega,g$. Under these conditions the dissipation of the system oscillator is modelled as displacement noise using the Brownian master equation.
	}
	\label{figDispNoise}
\end{figure}

As an illustrative example we consider a harmonic oscillator subject to displacement noise induced by a second, lossy oscillator. Although much simpler than the rigorous derivation of the Brownian motion master equation based on the Caldeira-Leggett model~\cite{BreuerPetruccione,CaldeiraLeggett}, this example and the resulting master equation already provide several identifying features of Brownian motion and illustrates the type of displacement noise that is common of processes such as recoil heating in AMO and levitated optomechanics~\cite{CiracPRA1992,GonzalezBallesteroPRA2019}. We consider a system of two harmonic oscillators of the same frequency $\Omega$, coupled at a rate $g$ (Fig.~\ref{figDispNoise}). Oscillator number $1$ forms the system, whereas oscillator number $2$ forms the bath. The latter is coupled to a thermal bath at temperature $T$ with a decay rate $\gamma \gg \Omega,g$. The Liouvillians for this system read in the Schr\"odinger Picture
\begin{equation}
    \Lv_S (*) = -\frac{i}{\hbar}\left[\hbar\Omega\hat{b}^\dagger_1\hat{b}_1, (*)\right],
\end{equation}
\begin{equation}
    \Lv_{\rm Int} (*) = -\frac{i}{\hbar}\left[\hbar g \hat{q}_1\hat{q}_2, (*)\right],
\end{equation}
\begin{multline}
    \Lv_B (*) = -\frac{i}{\hbar}\left[\hbar\Omega\hat{b}^\dagger_2\hat{b}_2, (*)\right]+
    \\
    +\gamma\bar{n}\mathfrak{D}_{\hat{b}^\dagger_2}[(*)]+\gamma(\bar{n}+1)\mathfrak{D}_{\hat{b}_2}[(*)],
\end{multline}
where $\hat{q}_j\equiv \hat{b}^\dagger_j+\hat{b}_j$ is the dimensionless position coordinate of oscillator $j$, $\bar{n} = (\exp[\hbar\Omega_2 \beta]-1)^{-1}$ is the Bose-Einstein distribution with $\beta \equiv (k_B T)^{-1}$ and $k_B$ the Boltzmann constant, and where we define the Lindblad dissipator~\cite{Lindblad1,Lindblad2GoriniKossakowskiSudarshan,tutorial2}
\begin{equation}\label{Lindbladdef}
    \mathfrak{D}_{\hat{A}}[(*)]\equiv\hat{A}(*)\hat{A}^\dagger-\frac{1}{2}\{\hat{A}^\dagger\hat{A},(*)\},
\end{equation}
with the curly brackets denoting the anticommutator.

The above Liouvillians clearly fulfill the hierarchy Eq.~\eqref{LiouvillianAE1} due to the assumption $\gamma \gg \Omega,g$, and thus we can use the expansion Eq.~\eqref{NZreducedBM}. The steady state of the bath, i.e., of oscillator 2, is a thermal state,
\begin{equation}
    \rho_B = Z^{-1}\exp(-\beta \hbar \Omega \hat{b}^\dagger_2\hat{b}_2),
\end{equation}
with $Z$ the partition function.
For this steady state it is straightforward to show that $\Py \Lv_{\rm Int} \Py = 0$ and thus we can cast Eq.~\eqref{NZreducedBM} in the simplified form
\begin{multline}\label{prov0}
    \dot{\rho}_S(t) \approx\Lv_S\rho_S(t)
    +\\+\text{Tr}_B \Lv_{\rm Int}\int_0^\infty d\tau e^{\Lv_B \tau} \Lv_{\rm Int} v(t).
\end{multline}
By repeated application of the interaction Liouvillian we can cast the second term in the above equation as
\begin{multline}\label{prov1}
    \text{Tr}_B \Lv_{\rm Int}\int_0^\infty d\tau e^{\Lv_B \tau} \Lv_{\rm Int} v(t) =
    \\
    =-\Big(F_1 \hat{q}_1\left[\hat{q}_1,\rho_S(t)\right]
    -
    F_2 \left[\hat{q}_1,\rho_S(t)\right]\hat{q}_1
    \Big)
\end{multline}
with the rates
\begin{equation}
    F_1 \equiv g^2\int_0^\infty d\tau \text{Tr}_B\left[\hat{q}_2\left(e^{\Lv_B\tau}\hat{q}_2\rho_B\right)\right],
\end{equation}
and
\begin{equation}
    F_2 \equiv g^2\int_0^\infty d\tau \text{Tr}_B\left[\hat{q}_2\left(e^{\Lv_B\tau}\rho_B\hat{q}_2\right)\right].
\end{equation}
The above expressions can be computed from the quantum regression theorem (see Appendix~\ref{appendix}), but also by using the identity $\rho_B\hat{b}_2 = e^{\hbar \beta \Omega}\hat{b}_2\rho_B$, which can be derived using the Baker-Campbell-Haussdorf formula. Using this identity one can compute 
\begin{equation}
    \Lv_B\left[
    \begin{array}{c}
         \rho_{B} \hat{b}_2  \\
         \hat{b}_2 \rho_{B} \\
         \rho_{B} \hat{b}_2^\dagger  \\
         \hat{b}_2^\dagger \rho_{B}
    \end{array}
    \right]
    =
    \left[
    \begin{array}{c}
         (i\Omega-\gamma/2)\rho_{B} \hat{b}_2  \\
         (i\Omega-\gamma/2)\hat{b}_2 \rho_{B} \\
         (-i\Omega-\gamma/2)\rho_{B} \hat{b}_2^\dagger  \\
         (-i\Omega-\gamma/2)\hat{b}_2^\dagger \rho_{B}
    \end{array}
    \right],
\end{equation}
from which the rates $F_{1,2}$ can be directly evaluated. We obtain
\begin{multline}
    F_1 = F_2^* = g^2\frac{-i\Omega + (2\bar{n}+1)\gamma/2}{\Omega^2 + (\gamma/2)^2}
    \\
    \approx
   \frac{2 g^2}{\gamma}(2\bar{n}+1) 
    \equiv\frac{\Gamma}{2},
\end{multline}
where in the second line we have used our assumption $\gamma \gg \Omega$. Introducing this result into Eqs.~\eqref{prov0} and \eqref{prov1} we obtain our final master equation
\begin{equation}\label{BrownianME}
    \dot{\rho}_S(t) \approx\Lv_S\rho_S(t)
    -\frac{\Gamma}{2} \left[\hat{q}_1,\left[\hat{q}_1,\rho_S(t)\right]\right].
\end{equation}
The dissipator we have obtained is often called ``momentum diffusion'' in the context of Brownian motion, or ``position delocalization'' as it produces a decay of the off-diagonal elements of the density matrix in position representation~\cite{Diosi2022}. This decoherence mechanism is typical of laser shot noise in optical tweezer traps for atoms or for dielectric nanoparticles~\cite{GonzalezBallesteroPRA2019}, and of collisions of gas molecules against a mechanical resonator~\cite{Diosi2022}. While in the former case this decoherence can be made subdominant by detuning the atom drive from the atomic transition~\cite{CiracPRA1992}, for nanoparticles it remains the main source of motional decoherence and a major obstacle on the road toward generating quantum motional states~\cite{RodaLlordesPRL2024,NeumeierPNAS2024}.

\section{Particular case 2: Bath decays faster than interaction rate (Born-Markov master equation, sideband cooling adiabatic elimination)}\label{SecBornMarkov}

Our next particular case focus on a paradigm of open quantum systems which is commonly referred to as the Born-Markov master equation. In the context of optomechanics~\cite{WilsonRaeNJP2008} and AMO~\cite{CiracPRA1992} sideband cooling (see example~\ref{exampleAEQO} below) it is also referred to as adiabatic elimination. We consider again a bath that decays on a fast timescale. In this case, however, the decay of the bath is not the fastest timescale of the problem: we allow the coherent timescales of both system and bath (typically their natural frequencies) to be equal or much faster. A paradigmatic example (discussed in Example~\ref{ExamplePurcell} below) is an atom transition coupled to an optical cavity whose decay rate is much smaller than the frequencies of both systems (typically $\sim 10^{15}$ Hz) but much larger than the atom-cavity coupling rate. Formally, we consider the following Liouvillian structure:
\begin{equation}\label{LiouvAEBM}
    \Lv(t) = \xi^2 \Lv_{B} + \xi\Lv_{\rm Int}(\xi^2 t) + \Lv_S(\xi^2 t),
\end{equation}
with $\xi \gg 1$ a large expansion parameter, which again quantifies the dissipative part of the bath Liouvillian, i.e. $\min(\text{Re}[-\text{Eigenvalues}[\Lv_B]])\propto \xi^2$ (except for $\lambda_j=0$, see Sec.~\ref{SecBrownian}).
Although the above Liouvillian can apply to other scenarios, it is usually better understood assuming it describes a Liouvillian in the interaction picture. In such scenario, the terms $\Lv_B$ and $\Lv_S$ would only contain dissipative terms, and the time dependence on both $\Lv_S$ and $\Lv_{\rm Int}$ would stem from the transformation to the interaction picture, i.e., it would usually take the form of a sum of exponentials $e^{i(\omega_S\pm\omega_S') t}$ and $e^{i(\omega_S\pm\omega_B) t}$ respectively, with $\omega_S$ and $\omega_B$ typical coherent evolution frequencies
of system and bath. In this example the dissipative part of the Liouvillian $\Lv_B$ is assumed time-independent in the interaction picture, a common situation in typical systems (see examples below). The following perturbative expansion thus allows system and bath coherent frequencies to be arbitrarily large -- hence our explicit inclusion of the perturbation parameter as $\xi^2t$ in the time dependence of system and interaction Liouvillians --, but still assumes the bath dissipation to be much faster than (i) any dissipative timescale in $\Lv_S$ and (ii) the timescale at which system and bath interact, i.e., any timescale in $\Lv_{\rm Int}$. For more insight on the physical picture, we address the reader to the examples below.

Since, by assumption, the bath Liouvillian is time-independent and much larger than the system Liouvillian, we can apply Eq.~\eqref{NakajimasimplifiedTime} or, equivalently, the simplified version Eq.~\eqref{NZreducedAM}. Assuming for simplicity that $\Py\Lv_{\rm Int}\Py = 0$  (although this is not necessary), and changing integration variable in  Eq.~\eqref{NakajimasimplifiedTime} from $\tau$ to $\xi^2(t-\tau)$, we can write this equation as
\begin{multline}
    \dot{\rho}_S(t) =\Lv_S(t)\rho_S(t) +\text{Tr}_B\Lv_{\rm Int}(\xi^2 t)\int_0^{\xi^2t} d\tau
    \\
    \times e^{\Lv_B \tau}\Lv_{\rm Int} (\xi^2t- \tau) v(t-\tau/\xi^2) + \mathcal{O}(\xi).
\end{multline}
We now take the limit $\xi \to \infty$. This results in an effective Markov approximation, as it allows us to drop the dependence of $v$ with $\tau$ and to extend the upper integration limit to infinity. Moreover, it is customary in this step to take, when possible, a rotating wave approximation~\cite{WilsonRaeNJP2008}. 
As we will show in the example below, this approximation consists in neglecting, in the product of the two interaction Liouvillians $\sim \Lv_{\rm Int}(\xi^2 t) (...) \Lv_{\rm Int}(\xi^2 t-\tau)$ , all the terms that oscillate rapidly in the variable $t$. In most cases, and in analogy to the elimination of ``fast time'' dependence in classical equations~\cite{VanKampen}, this makes the equation independent on $\xi^2t$, as one only retains terms in the product $\sim \Lv_{\rm Int}(\xi^2 t) (...) \Lv_{\rm Int}(\xi^2 t-\tau)$ whose $t-$dependencies cancel each other (see an example in the section below). We remark that, precisely because such dependencies can cancel each other, the dependence with $\tau$ might become relevant and thus it is in general not correct to
approximate $\Lv_{\rm Int}(\xi^2t-\tau) \approx \Lv_{\rm Int}(\xi^2 t)$ even in the large $\xi$ limit. This is the core difference with respect to the Brownian master equation in the previous section: in the Brownian case the final
effective dynamics only depend on steady-state bath expectation values and in the present case they depend on bath two-time
correlation functions.
After the rotating wave approximation, one finds the following Born-Markov or adiabatic elimination master equation~\cite{steck2007quantum,BreuerPetruccione},
\begin{multline}\label{finalAEOM}
    \dot{\rho}_S(t) =\Lv_S(t)\rho_S(t) +\text{Tr}_B\Lv_{\rm Int}(\xi^2 t)
    \\
    \times\int_0^{\infty} d\tau e^{\Lv_B \tau}\Lv_{\rm Int} (\xi^2t- \tau) v(t)\Big\vert_{\rm RWA}.
\end{multline}
As a final note, we remark that the rotating wave approximation is not in general only a matter of simplifying the equation. Indeed, it can be shown that without it the resulting effective dynamics are not guaranteed to be consistent with quantum mechanics, specifically the density matrix is not guaranteed to be positive semidefinite (this is shown in most derivations e.g. Ref.~\cite{BreuerPetruccione}, for a more in-depth picture see Ref.~\cite{Fleming_2010,tutorial5}). This also applies to the Brownian equation derived in the Sec.~\ref{SecBrownian}.
Note that if the rotating wave approximation is not taken the master equation, known in this case as Bloch-Redfield equation~\cite{REDFIELD19651}, can still be made quantum-consistent by other techniques (see e.g. Ref.~\cite{RedfieldEM} and references therein) and has been a powerful tool to explore the physics of
light harvesting complexes~\cite{JeskeJCP2015} or the vibrational relaxation of complex molecules~\cite{FigueiridoJCP92}, among others.

\subsection{Example: cavity cooling of a mechanical resonator}\label{exampleAEQO}

\begin{figure}[t]
	\centering
	\includegraphics[width=0.8\linewidth]{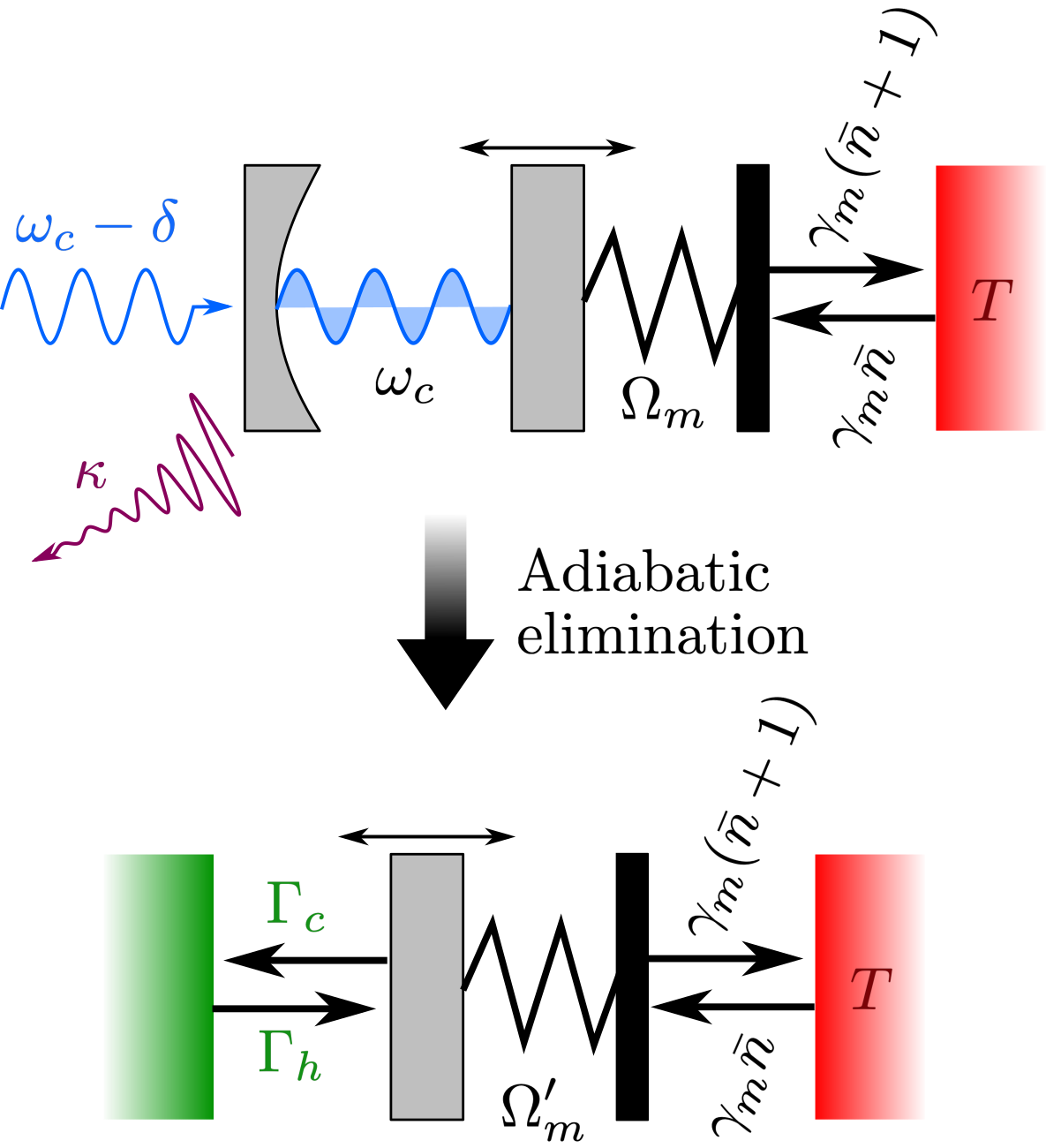}
	\caption{
	Under certain conditions, an optical cavity mode can be adiabatically eliminated and act as an effective environment for the mechanical motion, inducing cooling and heating rates $\Gamma_c$ and $\Gamma_h$. By tuning the parameters one can fix $\Gamma_c \gg \Gamma_h$, resulting in cavity cooling of the mechanical mode. 
	}
	\label{figCavityCooling}
\end{figure}

As a first relevant example we consider the optomechanical system depicted in Fig.~\eqref{figCavityCooling}. A mechanical mode (the system) with frequency $\Omega_m$ is coupled via radiation pressure to a cavity (the bath) with frequency $\omega_c$, which is coherently driven by a laser detuned by $\delta$. Both degrees of freedom experience dissipation through their coupling to thermal reservoirs. Specifically, the mechanical mode experiences absorption and decay at rates $\gamma_m\bar{n}$ and $\gamma_m(\bar{n}+1)$, with $\gamma_m$ the mechanical linewidth and $\bar{n}$ the average (Bose-Einstein) occupation of the reservoir at frequency $\Omega_m$ and temperature $T$. Due to its high (e.g., optical) frequency, the cavity experiences only decay at a rate $\kappa$, as its reservoir has near-zero occupation at such frequencies. 
In a frame rotating at the cavity driving frequency and after linearization of the optomechanical interaction~\cite{AspelmeyerRMP2014}, the Liouvillians of this problem take the form
\begin{equation}
    \Lv_S[(*)] = -i\left[\Omega_m\hat{b}^\dagger\hat{b},(*)\right] + \gamma_m \mathcal{D}_S[(*)],
\end{equation}
\begin{equation}
    \Lv_B[(*)] = -i\left[\delta\hat{a}^\dagger\hat{a},(*)\right] + \kappa \mathcal{D}_B[(*)],
\end{equation}
\begin{equation}
    \Lv_{\rm Int}[(*)] = -i\left[(g\hat{a}+g^*\hat{a}^\dagger)(\hat{b}+\hat{b}^\dagger),(*)\right],
\end{equation}
with $g$ the linearized optomechanical coupling, and with mechanical and optical dissipative terms given by
\begin{equation}
    \mathcal{D}_S[(*)] = \bar{n}\mathfrak{D}_{\hat{b}^\dagger}[(*)]+(\bar{n}+1)\mathfrak{D}_{\hat{b}}[(*)],
\end{equation}
and
\begin{equation}
    \mathcal{D}_B[(*)] = \mathfrak{D}_{\hat{a}}[(*)],
\end{equation}
in terms of the Lindblad dissipator defined in Eq.~\eqref{Lindbladdef}.
For convenience, we will work in a frame where the mechanical degrees of freedom rotate at the mechanical frequency (i.e., a frame defined by the time-dependent unitary $\exp[-i\Omega_m t\hat{b}^\dagger\hat{b}])$, where the system and interaction Liouvillians reduce to
\begin{equation}
    \Lv_S[(*)] = \gamma_m \mathcal{D}_S[(*)],
\end{equation}
\begin{multline}
    \Lv_{\rm Int}[(*),t] = 
    \\ -i\big[(g\hat{a}+\text{H.c.})(\hat{b}e^{-i\Omega_m t}+\text{H.c.}),(*)\big].
\end{multline}

It is well known in optomechanics~\cite{AspelmeyerRMP2014} that, under certain conditions, the cavity mode can be used to cool the mechanical mode well below the temperature of its surrounding environment, $T$. One regime where this occurs is the regime $\kappa,\Omega_m \gg \gamma_m,g$, which will be our regime of interest. Since the cavity decay $\kappa$ is very fast, the full dynamics must be well approximated by their projection onto the cavity steady state, namely its vacuum state. Hence, the cavity can be adiabatically eliminated using Eq.~\eqref{finalAEOM}. Note that this procedure can be different from the ``AMO adiabatic elimination'' studied in Sec.~\ref{SecAMOAE}, which similarly to the Brownian master equation in Sec.~\ref{SecBrownian} assumes the bath to be faster than all other timescales. 

The steady state of the cavity defining the projector is the vacuum state of the cavity mode $\hat{a}$, i.e., $\rho_B = \vert 0\rangle\langle 0 \vert$. It is straightforward to prove that in this case
\begin{equation}
    \Py\Lv_{\rm Int}\Py = 0.
\end{equation}
Thus, Eq.~\eqref{finalAEOM} for this problem reads
\begin{multline}\label{initialexampleEq}
    \dot{\rho}_S(t) =\gamma_m\mathcal{D}_S[\rho_S] + \text{Tr}_B\Lv_{\rm Int} (t)
    \\
    \times\int_0^{\infty} d\tau e^{\Lv_B \tau}\Lv_{\rm Int} (t- \tau) \rho_B\otimes\rho_S(t)\Big\vert_{\rm RWA}.
\end{multline}
where $\rho_S(t)$ is in this case the reduced density matrix of the mechanical mode.
Let us develop the above expression to derive the final master equation for the motion.

We first aim at performing the rotating wave approximation. To do that, we cast the interaction Liouvillian as
\begin{equation}
    \Lv_{\rm Int}(t) \equiv \sum_{\lambda=\pm}e^{i\lambda\Omega_mt}\Lv_\lambda,
\end{equation}
with time-independent superoperators
\begin{multline}
    \Lv_+[(*)] \equiv-i\left[\hat{b}^\dagger\otimes(g\hat{a}+\text{H.c.}),(*)\right]
    \\
    \equiv -i\left[\hat{b}_+\otimes\hat{Q}_a,(*)\right],
\end{multline}
\begin{multline}
    \Lv_-[(*)] \equiv-i\left[\hat{b}\otimes(g\hat{a}+\text{H.c.}),(*)\right]
    \\
    \equiv -i\left[\hat{b}_-\otimes\hat{Q}_a,(*)\right],
\end{multline}
where the notation $\hat{b}_\pm$ is used for convenience in the manipulation of the expressions. After introducing the above expressions into the second term of Eq.~\eqref{initialexampleEq} we obtain four terms: two of them with time dependence $\propto \exp[\tau(\Lv_B\pm i\Omega_m)]$, and two of them with time dependence $\propto \exp[\tau(\Lv_B\pm i\Omega_m)] \times \exp[\pm i 2\Omega_m t]$. Since by assumption $\Omega_m\gg \gamma_m,g$, the latter oscillate very fast as compared to the the system density matrix in the interaction picture (i.e., to the system lifetime $\gamma_m^{-1}$) or to the system-bath interaction timescales. Thus, these terms can be discarded under a rotating wave approximation. After this approximation we can write Eq.~\eqref{initialexampleEq} in compact form as
\begin{multline}\label{initialexampleEq2}
    \dot{\rho}_S(t) \approx\gamma_m\mathcal{D}_S[\rho_S] +\sum_{\lambda=\pm} \text{Tr}_B\Lv_{-\lambda} 
    \\
    \times\int_0^{\infty} d\tau e^{\tau(\Lv_B -i\lambda\Omega_m)}\Lv_{\lambda}   \rho_B\otimes\rho_S(t).
\end{multline}

The second step is to apply all the superoperators inside the integral to the projected density matrix $\rho_B\otimes\rho_S(t)$. The following three useful identities are straightforward to demonstrate:
\begin{multline}
    \Lv_{\lambda}   \rho_B\otimes\rho_S(t) = -i(g^*\hat{b}_\lambda\rho_S(t) \otimes\vert 1\rangle \langle 0 \vert
    \\
    -g\rho_s(t)\hat{b}_\lambda \otimes \vert 0\rangle\langle 1\vert),
\end{multline}
\begin{equation}\label{pprovv1}
    \Lv_B \vert 1\rangle\langle 0\vert = \left(-i\delta-\frac{\kappa}{2}\right)\vert 1\rangle\langle 0\vert,
\end{equation}
\begin{equation}\label{pprovv2}
    \Lv_B \vert 0\rangle\langle 1\vert = \left(i\delta-\frac{\kappa}{2}\right)\vert 0\rangle\langle 1\vert,
\end{equation}
where $\vert 1\rangle$ is the $n=1$ Fock state of the cavity mode.
Using these identities, Eq.~\eqref{initialexampleEq2} reduces to
\begin{multline}\label{initialexampleEq3}
    \dot{\rho}_S(t) =\gamma_m\mathcal{D}_S[\rho]
    \\-\vert g \vert^2\sum_{\lambda=\pm} E_\lambda(\Omega_m)\left[\hat{b}_{-\lambda}\hat{b}_\lambda\rho-\hat{b}_\lambda\rho\hat{b}_{-\lambda}\right]
    \\
    +\vert g \vert^2\sum_{\lambda=\pm}E_\lambda^*(-\Omega_m)\left[\hat{b}_{-\lambda}\rho\hat{b}_\lambda-\rho\hat{b}_\lambda\hat{b}_{-\lambda}\right]
    .
\end{multline}
where we have defined
\begin{multline}\label{PSDprov}
    E_\lambda(\omega) \equiv \int_0^\infty d\tau e^{\tau[-i(\delta+\lambda\omega)-\kappa/2]} =
    \\
    =\frac{1}{i(\delta+\lambda\omega)+\kappa/2}.
\end{multline}
Note that in the literature the above rates are often expressed in terms of the power spectral densities of the bath, $S(\omega)=(\pi)^{-1}\vert g\vert^2 E_-(\omega)=(\pi)^{-1}\vert g\vert^2 E_+(-\omega)$ \cite{BreuerPetruccione} (see also Appendix~\ref{appendix}).
Finally, by rearranging the terms and using the identity $E_\pm(\omega) = E_\mp(-\omega)$, we can cast the above equation into Lindblad form,
\begin{multline}\label{initialexampleEq4}
    \dot{\rho}_S(t) =\gamma_m\mathcal{D}_S[\rho]-i\left[\Delta_m\hat{b}^\dagger\hat{b},\rho\right]
    \\+\Gamma_h\mathfrak{D}_{\hat{b}^\dagger}[\rho]+\Gamma_c\mathfrak{D}_{\hat{b}}[\rho]
    \\
    =
    -i\left[\Delta_m\hat{b}^\dagger\hat{b},\rho\right] + \left(\gamma_m\bar{n}+\Gamma_h\right)\mathfrak{D}_{\hat{b}^\dagger}[\rho]
    \\
    +
    \left(\gamma_m(1+\bar{n})+\Gamma_c\right)\mathfrak{D}_{\hat{b}}[\rho],
\end{multline}
The adiabatically eliminated cavity thus induces, first, a frequency shift of the mechanical mode given by
\begin{equation}
    \Delta_m \equiv \vert g\vert^2\text{Im}\left[E_+(\Omega_m)+E_-(\Omega_m)\right].
\end{equation}
Second, it introduces additional heating and cooling rates given by
\begin{multline}
    \Gamma_h \equiv 2\vert g\vert^2\text{Re}[E_+(\Omega_m)]=
    \\=\frac{\vert g\vert^2\kappa}{(\delta+\Omega_m)^2+(\kappa/2)^2}
\end{multline}
\begin{multline}
    \Gamma_c \equiv 2\vert g\vert^2\text{Re}[E_-(\Omega_m)]=
    \\=\frac{\vert g\vert^2\kappa}{(\delta-\Omega_m)^2+(\kappa/2)^2}.
\end{multline}
This completes the adiabatic elimination.

Let us briefly discuss the physical implications of the above expression. 
The steady-state occupation of the mechanical mode can be easily derived from the master equation Eq.~\eqref{initialexampleEq4},
\begin{equation}
    \langle \hat{b}^\dagger\hat{b}\rangle_{\rm ss} = \frac{\gamma_m \bar{n} + \Gamma_h}{\gamma_m+\Gamma_c-\Gamma_h}.
\end{equation}
In the absence of the cavity ($g=0$) this occupation reduces to $\langle\hat{b}^\dagger\hat{b}\rangle_{\rm ss}=\bar{n}$, i.e., the system reaches thermal equilibrium with its reservoir. For efficient cooling and minimal heating of the mechanical motion, the first condition is that the cavity driving is blue-detuned with respect to the mechanical mode by exactly the mechanical frequency, i.e., $\delta = \Omega_m$. In these conditions one has
\begin{equation}
    \Gamma_c = \frac{4\vert g \vert^2}{\kappa}
\end{equation}
and
\begin{equation}
    \Gamma_h = \frac{\vert g\vert^2\kappa}{4\Omega_m^2+(\kappa/2)^2}.
\end{equation}
The second well-known condition for efficient cooling is to work in the ``resolved sideband regime''
\begin{equation}
    \kappa \ll \Omega_m,
\end{equation}
where one has $\Gamma_h \approx \Gamma_c [\kappa/(4\Omega_m)]^{2} \ll \Gamma_c$. In this limit the occupation reduces to
\begin{equation}
    \langle \hat{b}^\dagger\hat{b}\rangle_{\rm ss} \approx \frac{ \bar{n} }{1+C},
\end{equation}
where we have defined the cooperativity
\begin{equation}
    C\equiv \frac{\Gamma_c}{\gamma_m} =\frac{4\vert g\vert^2}{\kappa\gamma_m}.
\end{equation}
The final condition for efficient optomechanical cooling is to work in the high-cooperativity regime $C\gg 1$, where one can reach occupations $\langle \hat{b}^\dagger\hat{b}\rangle_{\rm ss} \ll \bar{n}$, and even below unity (ground-state cooling). 

\subsection{Example 2: Purcell effect}\label{ExamplePurcell}

\begin{figure}[t]
	\centering
	\includegraphics[width=0.7\linewidth]{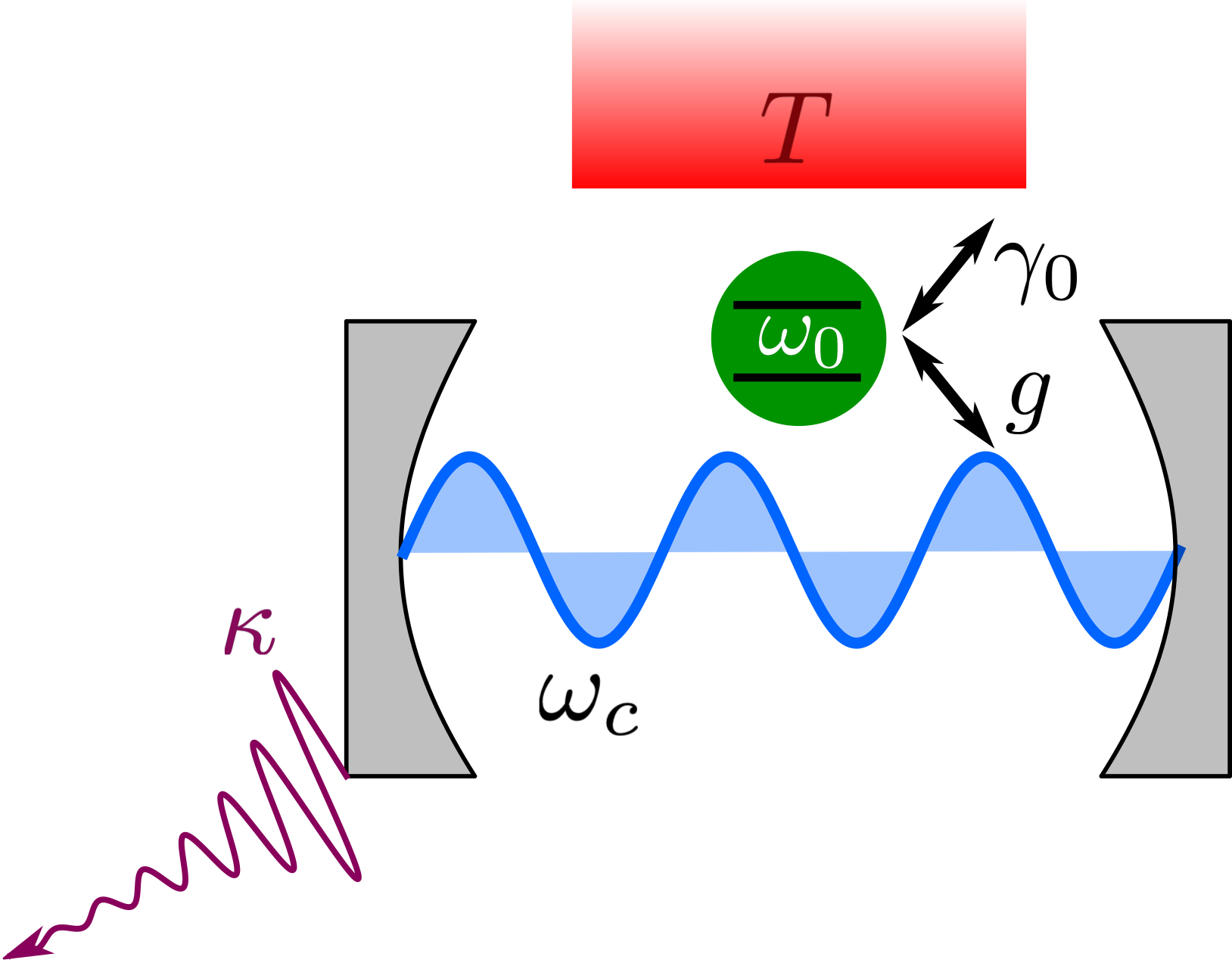}
	\caption{
	A two-level system with transition frequency $\omega_0$ is coupled at a rate $g$ to a cavity with frequency $\omega_c$ and decay rate $\kappa$, and at a rate $\gamma_0$ to a thermal reservoir. In the relevant regime $\omega_0,\omega_c \gg \kappa \gg g,\gamma_0$ the cavity can be traced out and the qubit is modelled by a Born-Markov master equation. The main effect of the cavity is a modification of the qubit decay rate (Purcell effect).
	}
	\label{figPurcell}
\end{figure}

As a second paradigmatic example in quantum optics, we consider the Purcell effect, namely the modification of a quantum emitter's lifetime when placed inside an optical cavity within the weak coupling regime~\cite{Purcell,Purcell2,steck2007quantum,NavarreteQOCourse,CohenAtomPhotonInteractions,steck2007quantum}. Specifically we consider the system of Fig.~\ref{figPurcell}: a two-level system with frequency $\omega_0$ is coupled to a thermal reservoir with decay rate $\gamma_0$ and to an optical cavity with frequency $\omega_c$ and linewidth $\kappa$ at a rate $g$. In the interaction picture, the Liouvillians are given by
\begin{equation}
    \Lv_S^{(i)}[(*)] = \gamma_0 \mathcal{D}_S^{(i)}[(*)],
\end{equation}
\begin{equation}
    \Lv_B^{(i)}[(*)] = \kappa \mathcal{D}_B^{(i)}[(*)],
\end{equation}
\begin{multline}\label{LintJaynesCummings}
    \Lv_{\rm Int}^{(i)}[(*)] = 
    \\-i\left[ge^{-i\delta t}\hat{a}^\dagger\hat{\sigma}_-+g^*e^{i\delta t}\hat{a}\hat{\sigma}_+,(*)\right],
\end{multline}
with
\begin{equation}
    \mathcal{D}_S[(*)] = \bar{n}\mathfrak{D}_{\hat{\sigma}^\dagger}[(*)]+(\bar{n}+1)\mathfrak{D}_{\hat{\sigma}}[(*)],
\end{equation}
and
\begin{equation}
    \mathcal{D}_B[(*)] = \mathfrak{D}_{\hat{a}}[(*)],
\end{equation}
where $\delta\equiv\omega_0-\omega_c$ and $\hat{\sigma}_- =\hat{\sigma}_+^\dagger \equiv \vert g\rangle \langle e \vert$ is the qubit lowering operator, with $\vert g \rangle$ and $\vert e \rangle $ the qubit ground and excited state respectively. The interaction Liouvillian Eq.~\eqref{LintJaynesCummings} has the usual Jaynes-Cummings form~\cite{NavarreteQOCourse,steck2007quantum} where the electric dipole interaction has been simplified by a rotating wave approximation.

This example is very similar to the previous one in Sec.~\ref{exampleAEQO}. In this case, however, the oscillation frequencies of the interaction Liouvillian are given by a detuning $\delta$ that can be arbitrarily small, a situation where the rotating wave approximation would not be valid. Note that this is not in conflict with the hierarchy Eq.~\eqref{LiouvAEBM}. To account for all possible regimes of $\delta$ we use the version of Eq.~\eqref{finalAEOM} without rotating wave approximation,
\begin{multline}\label{initialexampleEq2IP}
    \dot{\rho}_S^{(i)}(t) =\gamma_0\mathcal{D}_S^{(i)}[\rho_S] + \text{Tr}_B\Lv_{\rm Int}^{(i)} ( t)
    \\
    \times\int_0^{\infty} d\tau e^{\Lv_B^{(i)} \tau}\Lv_{\rm Int}^{(i)} ( t- \tau) \rho_B\otimes\rho_S^{(i)}(t),
\end{multline}
where we have used that for this case and for the bath steady state (i.e., the cavity vacuum) one has $\Py \Lv_{\rm Int}\Py = 0$. We solve this equation following identical steps as in the previous example. First, we derive similar expressions to \eqref{pprovv1}-\eqref{pprovv2} for this case,
\begin{equation}
    \Lv_B^{(i)} \left[
    \begin{array}{c}
         \vert 0 \rangle \langle 1 \vert  \\
         \vert 1 \rangle \langle 0 \vert 
    \end{array}
    \right] = -\frac{\kappa}{2}\left[
    \begin{array}{c}
         \vert 0 \rangle \langle 1 \vert  \\
         \vert 1 \rangle \langle 0 \vert 
    \end{array}
    \right],
\end{equation}
with $\vert 0 \rangle$ and $\vert 1\rangle$ the vacuum and single-photon states of the cavity respectively.
Then, using this identity, by sequential application of each operator we obtain 
\begin{multline}
    \text{Tr}_B \Lv_{\rm Int}^{(i)} (t) \int_0^\infty d\tau e^{\Lv_B^{(i)}\tau}\Lv_{\rm Int}^{(i)} (t-\tau)\rho_B^{(i)}\otimes\rho_S^{(i)}(t)
    \\
    =
    \vert g \vert^2 I_-\left[\hat{\sigma}_-,\rho_S(t)\hat{\sigma}_+\right]
    -
    \vert g \vert^2 I_+\left[\hat{\sigma}_+,\hat{\sigma}_-\rho_S(t)\right],
\end{multline}
with
\begin{equation}
    I_\pm \equiv \int_0^\infty d\tau e^{(\pm i\delta-\kappa/2)\tau}
    =
    \frac{\pm i\delta + \kappa/2}{\delta^2 + (\kappa/2)^2}.
\end{equation}
Splitting these quantities into real and imaginary parts, rearranging the terms and transforming back to the Schr\"odinger picture we find the final master equation
\begin{multline}
     \dot{\rho}_S(t) =-\frac{i}{\hbar}\left[\hbar(\omega_0-\Delta_0)\hat{\sigma}_+\hat{\sigma}_-,\rho_S(t)\right]+
     \\
     +\gamma_0\mathcal{D}_S[\rho_S] +\Gamma \mathfrak{D}_{\hat{\sigma}}[\rho_S(t)],
\end{multline}
i.e., the effect of the cavity is to induce a qubit frequency shift
\begin{equation}
    \Delta_0\equiv\frac{\delta\vert g \vert^2}{\delta^2+(\kappa/2)^2},
\end{equation}
and an additional qubit decay rate
\begin{equation}
    \Gamma \equiv \frac{\kappa\vert g \vert^2}{\delta^2+(\kappa/2)^2}.
\end{equation}
The cavity-induced modification of the qubit emission lifetime is called Purcell effect and is a core result of cavity QED.

\section{Particular case 3: Lossless but rapidly oscillating bath (AMO adiabatic elimination)}\label{SecAMOAE}

The last particular case has the particularity that the bath does not decay fast or even at all (i.e., the bath can be a closed system) but all its natural frequencies are much larger than the system frequencies and system-bath coupling rates. Typically this situation is addressed via a Schrieffer-Wolff transformation followed by a suitable truncation (See e.g. complement B.I. in Ref.~\cite{CohenAtomPhotonInteractions} and references therein), but in principle it can also be addressed using the projector approach~\cite{KesslerPRA2012}. We consider a Liouvillian hierarchy of very similar form of Sec.~\ref{SecBrownian},
\begin{equation}\label{LiouvillianAE1bb}
    \Lv (t) = \xi^2\Lv_B +\xi \Lv_{\rm Int} + \Lv_S,
\end{equation}
where for simplicity we assume all terms to be time-independent and where we have introduced a large parameter $\xi \gg 1$ quantifying the fast bath timescale. \textcolor{blue}{}. Since the bath Liouvillian is the largest energy scale in the system we can use as a starting point Eq.~\eqref{NZreducedAM} (we assume $\Py\mathcal{L}_{\rm Int}\Py$ = 0)
\begin{multline}\label{NZreducedAM2}
    \dot{\rho}_S(t) =\Lv_S\rho_S(t)+
    \\
    + \text{Tr}_B \Lv_{\rm Int}\int_0^t d\tau e^{\Lv_B \tau} \Lv_{\rm Int} v(t-\tau) .
\end{multline}
The key difference in this situation is that the bath Liouvillian is assumed purely conservative or dominated by a conservative part, such that $\text{Re}[\text{eigenvalues}[\Lv_B]] \ll \min(\text{Im}[\text{eigenvalues}[\Lv_B]]) \sim \xi^2$. In this situation the bath does not show a fast decay and the Born-Markov approximation used in Sec.~\ref{SecBornMarkov} cannot be applied in a straightforward way. However, one can still find an approximate effective dynamical equation that is time-local (i.e., Markovian) by exploiting the assumption that the bath coherent evolution is much faster than all other timescales.

To exploit such assumptions we can use the  eigenvectors defined in Eqs.~\eqref{righteigenvecs} and~\eqref{lefteigenvecs} to write Eq.~\eqref{NZreducedAM2} as
\begin{multline}\label{NZreducedAM3}
    \dot{\rho}_S(t) =\Lv_S\rho_S(t)+\sum_{j}\text{Tr}_B \int_0^t d\tau e^{\lambda_j\tau }
    \\
     \Lv_{\rm Int} \vert \mathbf{r}_j ) (\mathbf{l}_j\vert  \Lv_{\rm Int}\rho_B\otimes \rho_S(t-\tau) .
\end{multline}
Note that a ``steady state'' $\rho_B$ in the sense of Eq.~\eqref{steadystatechoice} can always be defined even if the bath does not decay. Assuming for simplicity that the system is finite, the system density matrix can also in general be formally written as a sum over different exponentials (by e.g. Fourier transform)~\cite{vectorization1,vectorization2}, 
\begin{equation}
    \rho_S(t) = \sum_l e^{\Lambda_l t} \rho_{S,l}.
\end{equation}
For an unperturbed system, i.e., if $\Lv_{\rm Int} =0$, the frequencies $\Lambda_l$ correspond to the eigenvalues of the system Liouvillian $\Lv_S$. When $\Lv_{\rm Int} \ne 0$, these frequencies are modified by the interaction with the bath. In the perturbative regime where the system-bath interactions are weak, the system frequencies are perturbed by at most~\cite{BreuerPetruccione,CohenAtomPhotonInteractions}
\begin{equation}
    \Lambda_l' = \Lambda_l + \mathcal{O}(\epsilon^2),
\end{equation}
where $\epsilon \ll 1$ is a small parameter quantifying weak system-bath coupling, see Sec.~\ref{subsecWC}.
Introducing this into Eq.~\eqref{NZreducedAM3} we obtain
\begin{multline}\label{NZreducedAM4}
    \dot{\rho}_S(t) =
    \\
    \Lv_S\rho_S(t)+ \text{Tr}_B \Lv_{\rm Int}\sum_{jl}\int_0^t d\tau e^{i\vert\eta_j\vert \xi^2\tau }
    \\
     \vert \mathbf{r}_j ) (\mathbf{l}_j\vert  \Lv_{\rm Int}\rho_B\otimes e^{[\Lambda_l+ \mathcal{O}(\epsilon^2)] (t-\tau)} \rho_{S,l} ,
\end{multline}
where we have explicitly written the bath Liouvillian eigenvalues as $\lambda_j \sim i\vert \eta_j\vert \xi^2$ with $\eta_j \sim \Lambda_l$ and neglected the small real parts of $\lambda_j$ that do not affect the validity of this argument. From the above representation it is straigthforward to prove that,
\begin{multline}
    \int_0^t d\tau e^{[i\eta_j\xi^2 - \Lambda_l - \mathcal{O}(\epsilon^2)]\tau} \approx
    \\\int_0^t d\tau e^{i\eta_j\xi^2 \tau} + \mathcal{O}\left(\Lambda_l/\xi^{-2}\right).
\end{multline}
In other words, to lowest order in $\xi^{-1}$ we can drop the $\tau$ dependence of the system density matrix within the integral, effectively making it local in time. In the original notation of Eq.~\eqref{NZreducedAM2} we can write
\begin{multline}\label{NZreducedAM5}
    \dot{\rho}_S(t) \approx\Lv_S\rho_S(t)+\\+
     \text{Tr}_B \Lv_{\rm Int}\int_0^t d\tau e^{\Lv_B \tau} \Lv_{\rm Int} v(t) .
\end{multline}
Note that this expression is formally very similar to the Brownian master equation Eq.~\eqref{NZreducedBM}, but we have arrived to it following a different set of assumptions.
In this case, the time-local approximation Eq.~\eqref{NZreducedAM5} can be understood as a rapid-averaging effect where the slowly evolving system density matrix is averaged out by the rapid oscillations of the bath except at very short delays $\tau$. It can also be understood as the result of a weak coupling between two widely separated spectral regions of the Liouvillian, which therefore remain widely separated~\cite{CohenAtomPhotonInteractions}. A crucial difference between Eq.~\eqref{NZreducedAM5} and \eqref{NZreducedBM} is that in the former the upper integration limit is not taken to infinity as the non-decaying integrand does not allow for such approximation. Hence, instead of the irreversible system decays induced by the Brownian motion master equation Eq.~\eqref{NZreducedBM}, the solutions of Eq.~\eqref{NZreducedAM5} will (if $\Lv_S$ has no dissipative part) be purely periodic with a recurrence time $\sim (\eta_j\xi^2)^{-1}$.
We emphasize that for simplicity we have presented in this section only a short and intuitive version of the derivation, but a rigorous derivation can be found in the literature~\cite{KesslerPRA2012}.

\subsection{Example: Adiabatic elimination in a Lambda System}\label{SecLambdaSystem}

\begin{figure}[t]
	\centering
	\includegraphics[width=0.8\linewidth]{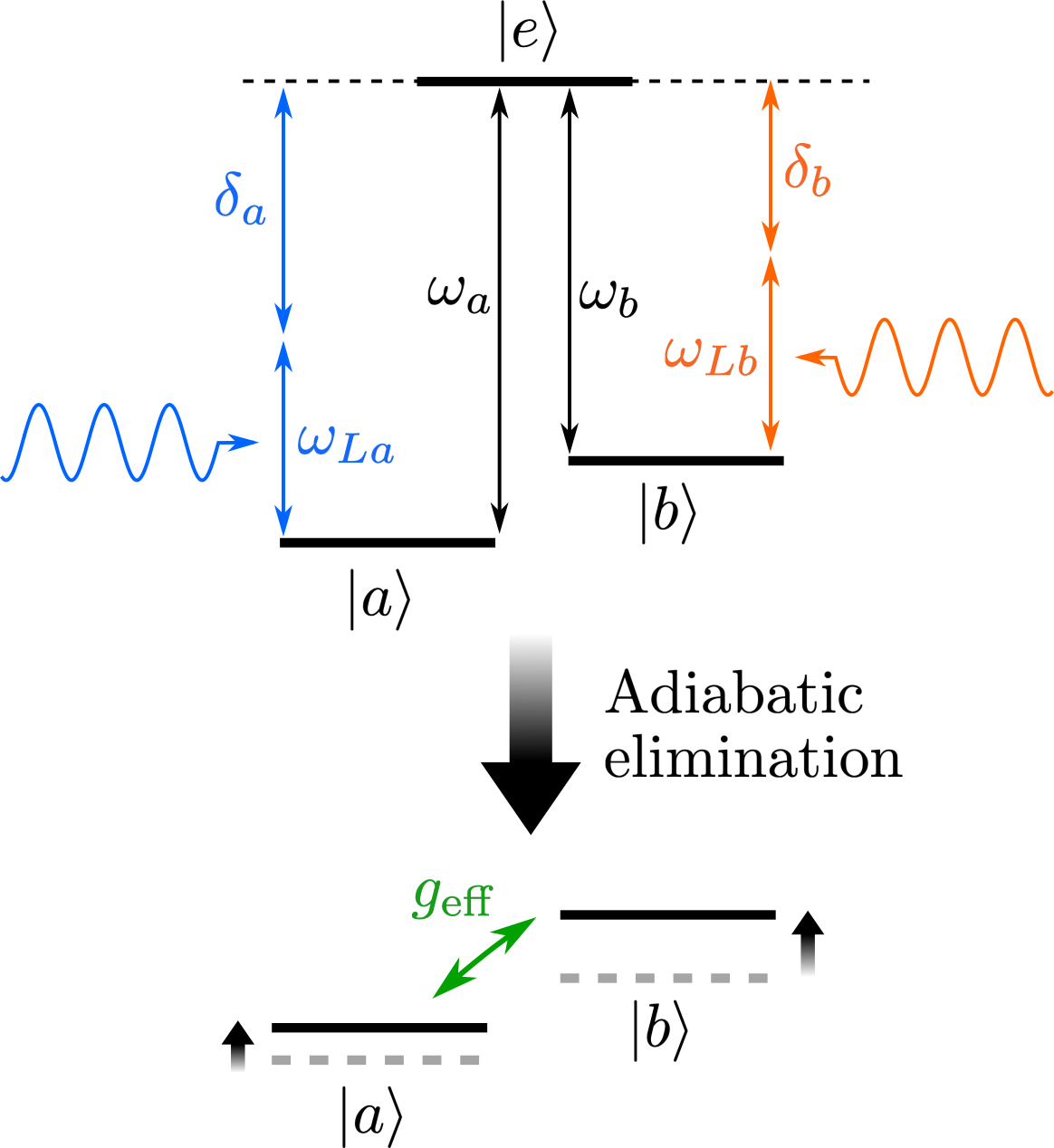}
	\caption{
	Example 1: the two transitions of a three-level system in a Lambda configuration are coherently driven. When the two drivings are far detuned, the excited state remains practically unpopulated, but can mediate transitions between the two ground states $\vert a\rangle$ and $\vert b\rangle$. In this limit one can obtain, via adiabatic elimination, an effective Hamiltonian for the ground-state manifold in which the effect of the drivings and the excited state is mapped into frequency shifts and an effective coupling term~\cite{BrionJPhysA2007}. 
	}
	\label{fig3LS}
\end{figure}

For this example we choose a canonical example of adiabatic elimination in atom quantum optics. The system we consider is a three-level system in a Lambda configuration, formed by two lower energy states $\vert a\rangle$ and $\vert b\rangle$ and an excited state $\vert e\rangle$ [see Fig.~\eqref{fig3LS}]. We follow the notation of Ref.~\cite{BrionJPhysA2007}, where this problem has been discussed in depth. We define the frequency of the transitions $\vert a\rangle\to\vert e\rangle$ and $\vert b\rangle \to \vert e\rangle$ as $\omega_a$ and $\omega_b$, respectively. A coherent drive is applied to each of these transitions, with respective rates $\Omega_{a}$ and $\Omega_b$ and respective frequencies $\omega_{La}$ and $\omega_{Lb}$. We assume the system experiences no dissipation. In the rotating wave approximation, the Hamiltonian of this system reads
\begin{multline}
    \hat{H}/\hbar = \omega_a\vert e\rangle\langle e\vert + (\omega_a-\omega_b)\vert b\rangle\langle b\vert + 
    \\
    +\sum_{j=a,b}\left(\frac{\Omega_j}{2}e^{-i\omega_{Lj}t}\vert e\rangle\langle j \vert+\text{H.c.}\right).
\end{multline}
Notice we take the level $\vert a\rangle$ as the origin of energies. We simplify the Hamiltonian by applying the unitary transformation
\begin{multline}
    \hat{U} = \exp\Big[i t\Big( (\omega_a-\omega_b-\delta/2)\vert b\rangle\langle b\vert +
    \\ +(\delta/2)\vert a \rangle\langle a\vert
    +(\omega_a-\Delta)\vert e\rangle\langle e\vert)\Big)\Big],
\end{multline}
where we define the detunings between each coherent drive and the corresponding transition frequencies, $\delta_j \equiv \omega_j-\omega_{Lj}$, their difference $\delta=\delta_a-\delta_b$, and the average detuning $\Delta \equiv (\delta_a+\delta_b)/2$. Under this transformation the Hamiltonian becomes time-independent,
\begin{multline}
    \hat{H}/\hbar = -\frac{\delta}{2}\vert a\rangle\langle a\vert
    +\frac{\delta}{2}\vert b\rangle\langle b\vert+\Delta
    \vert e\rangle\langle e\vert \\
    +\sum_{j=a,b}\left(\frac{\Omega_j}{2}\vert e\rangle\langle j \vert+\text{H.c.}\right).
\end{multline}

We now focus on the case where the excited level is far detuned with respect to the drivings, i.e., we assume
\begin{equation}\label{assumptionAE1}
    \vert\Delta\vert \gg \vert\delta\vert,\vert\Omega_{j}\vert.
\end{equation}
In this limit the level $\vert e\rangle$ does not play an active role in the dynamics, always remaining unpopulated. However, it can mediate virtual transitions between the levels $\vert a\rangle$ and $\vert b\rangle$, resulting in an effective coupling between them. Our goal is thus to adiabatically eliminate the level $\vert e\rangle$ and derive a reduced equation of motion in the subspace spanned by the states $\vert a\rangle$ and $\vert b\rangle$. We thus split the Hamiltonian into system, bath, and interaction part, $\hat{H} = \hat{H}_S + \hat{H}_B +\hat{V}$, where
\begin{equation}
    \hat{H}_S/\hbar = -\frac{\delta}{2}\vert a\rangle\langle a\vert
    +\frac{\delta}{2}\vert b\rangle\langle b\vert,
\end{equation}
\begin{equation}
    \hat{H}_B/\hbar = \Delta \vert e\rangle\langle e\vert,
\end{equation}
\begin{equation}\label{HI}
    \hat{V}/\hbar = \vert e\rangle\sum_{j=a,b}\frac{\Omega_j}{2}\langle j \vert +\text{H.c.}
\end{equation}
The corresponding Liouvillians are time-independent and given by
\begin{equation}
    \Lv_{\{S,B\}}(*) =-\frac{i}{\hbar}\left[\hat{H}_{\{S,B\}},(*)\right],
\end{equation}
\begin{equation}
    \Lv_{\rm Int} = -\frac{i}{\hbar}\left[\hat{V},(*)\right].
\end{equation}

A necessary preliminary step to use the formalism of this tutorial is to write the interaction Hamiltonian Eq.~\eqref{HI} in the form of a sum of tensor products of system and bath operators, as in Eq.~\eqref{V0000}. To do so, we \textit{artificially} enlarge our Hilbert space, i.e., assume
it is formed by two subsystems, a bath $E$ with ground state $\vert 0_e\rangle$ and a system $AB$ with ground state $\vert 0_s\rangle$, connected to the physical states  $\vert j \rangle$ ($j=a,b,e$) via ladder operators $\hat{j}$ and $\hat{j}^\dagger$ fulfilling
\begin{equation}
    \hat{j}\vert 0_s\rangle = \hat{j}^\dagger \vert j'\rangle = 0,
\end{equation}
\begin{equation}
    \hat{j}^\dagger\vert0_s\rangle = \vert j \rangle \hspace{0.3cm} ; \hspace{0.3cm}  \hat{j} \vert j'\rangle = \delta_{jj'}\vert0_s\rangle ,
\end{equation}
for $(j,j')\in (a,b)$, and
\begin{equation}
    \hat{e}\vert 0_e\rangle = \hat{e}^\dagger \vert e\rangle = 0,
\end{equation}
\begin{equation}
      \hat{e}^\dagger\vert 0_e\rangle = \vert e \rangle \hspace{0.3cm} ; \hspace{0.3cm} \hat{e} \vert e\rangle = \vert 0_e\rangle.
\end{equation}
These operators are artificial constructs which don't necessarily have a well-defined physical meaning (although products of two, four, etc of these operators do~\cite{SanchezBurilloPRA2020}). They are just helpful representations that allow us to compute the reduced dynamics. In terms of these operators the interaction Hamiltonian Eq.~\eqref{HI} can be cast in the desired form as
\begin{equation}\label{HIAE1ops}
    \hat{V}/\hbar = \hat{e}^\dagger \otimes \hat{S} +\hat{e} \otimes \hat{S}^\dagger,
\end{equation}
with a global system jump operator
\begin{equation}
    \hat{S} \equiv \sum_{j=a,b}\frac{\Omega_j}{2}\hat{j}.
\end{equation}
Using this representation, the partial trace over the bath degrees of freedom, namely the system $E$, reads
\begin{equation}
    \text{Tr}_B[(*)] = \langle e\vert (*)\vert e \rangle+\langle 0_e\vert (*)\vert 0_e \rangle.
\end{equation}
We define the projector as
\begin{multline}\label{defprojectorAE1}
    \Py[(*)] \equiv \rho_B \otimes \text{Tr}_B[(*)] \equiv \\\vert 0_e\rangle\langle 0_e \vert \otimes \text{Tr}_B[(*)],
\end{multline}
The choice of the bath state $\rho_B=\vert 0_e\rangle\langle 0_e\vert$ is physically motivated, as it is the only state in the subspace spanned by $\{\vert0_e\rangle,\vert e\rangle\}$ fulfilling the following conditions: first, the probability of the whole system to be in state $\vert e\rangle$ is zero, $\langle e \vert \rho_B \vert e\rangle =0$. This is consistent with our assumption that the excited state $\vert e\rangle$ is never populated during the dynamics. Second, the state $\rho_B$ is a stationary state of the bath Liouvillian, $\Lv_B(\vert 0_e\rangle\langle 0_e\vert)=0$.

We are now ready to proceed with the adiabatic elimination.
Due to the assumption Eq.~\eqref{assumptionAE1}, the Liouvillian of our system obeys the hierarchy of energy scales of the general Liouvillian Eq.~\eqref{LiouvillianAE1bb}, allowing to use the expansion Eq.~\eqref{NZreducedAM5}. Note that in this case the Markovian version Eq.~\eqref{NZreducedBM} cannot be used as the exponential $\exp(\mathcal{L}_B\tau)$ does not decay due to the absence of bath dissipation. Equation~\eqref{NZreducedAM5} for this example reads
\begin{multline}
     \dot{\rho}_S(t) \approx -\frac{i}{\hbar}\left[\hat{H}_S,\rho_S\right]-\frac{1}{\hbar^2}\int_0^t d\tau
    \\ 
    \text{Tr}_B\left[\hat{V},e^{\Lv_B \tau}\left[\hat{V},\rho_B\otimes\rho_S(t)\right]\right],
\end{multline}
where we have used the fact that the interaction Liouvillian fulfills $\Py\Lv_{\rm Int}\Py=0$.
We can compute the second line by introducing the explicit expression of $\hat{V}$, Eq.~\eqref{HIAE1ops}, and applying each superoperator sequentially. This requires to compute the action of the superoperator $\exp[\Lv_B\tau]$ on bath density matrices. Since the bath space $E$ has only dimension $2$, we need only to compute the action on the four matrices $\vert e\rangle\langle e\vert$, $\vert 0_e\rangle\langle e\vert$, $\vert e\rangle\langle 0_e\vert$, and $\vert 0_e\rangle\langle 0_e\vert$. From the definition of $\Lv_B$ it is straightforward to prove that
\begin{equation}
    \Lv_B \vert e\rangle\langle e\vert = \Lv_B \vert 0_e\rangle\langle 0_e\vert = 0,
\end{equation}
\begin{equation}
    \Lv_B \vert 0_e\rangle\langle e\vert = i\Delta \vert 0_e\rangle\langle e\vert
\end{equation}
\begin{equation}
     \Lv_B \vert e\rangle\langle 0_e\vert = -i\Delta \vert e\rangle\langle 0_e\vert,
\end{equation}
from which the identities
\begin{equation}
    e^{\Lv_B \tau} \vert e\rangle\langle e\vert = \vert e\rangle\langle e\vert,
\end{equation}
\begin{equation}
    e^{\Lv_B \tau} \vert 0_e\rangle\langle  0_e\vert = \vert  0_e\rangle\langle  0_e\vert,
\end{equation}
\begin{equation}
    e^{\Lv_B \tau} \vert 0_e\rangle\langle e\vert = e^{i\Delta \tau}\vert 0_e\rangle\langle e\vert,
\end{equation}
\begin{equation}
    e^{\Lv_B \tau} \vert e\rangle\langle 0_e\vert = e^{-i\Delta \tau}\vert e\rangle\langle 0_e\vert,
\end{equation}
immediately follow. Using these identities, we obtain the master equation
\begin{multline}
    \dot{\rho}_S(t) = -\frac{i}{\hbar}\left[\hat{H}_S,\rho_S\right] + 
    \\
    +\frac{i}{\Delta}\Big[-\rho_S\hat{S}^\dagger\hat{S}\left(1-e^{i\Delta t}\right)
    \\
    +\hat{S}^\dagger\hat{S}\rho_S\left(1-e^{-i\Delta t}\right)\\-2 i \sin(\Delta t)\hat{S}\rho_s\hat{S}^\dagger\Big].
\end{multline}
In this form, we are ready to dispose of the operator notation and rewrite the equation in terms of the original
states $\vert a\rangle$ and $\vert b \rangle$. The term proportional to $\hat{S}\rho_S\hat{S}^\dagger$ can be neglected as it does not act on the subspace spanned by $\vert a\rangle$ and $\vert b \rangle$ (or, equivalently, $\langle j\vert \hat{S}\rho_S\hat{S}^\dagger \vert k\rangle = 0$ for all $j,k=a,b$). Furthermore, the terms oscillating at frequency $\Delta$ can be discarded under a rotating wave approximation. This approximation is valid provided that $\vert \Delta\vert$ oscillates much faster than the oscillation frequencies of $\hat{S}^\dagger\hat{S}$, i.e., $\vert \Delta\vert \gg \vert \delta\vert$, and provided that $\vert\Delta\vert$ is much larger than the oscillation amplitude, namely $\vert \Delta\vert \gg \Omega_{a,b}^2/\vert\Delta \vert$. Both these conditions are true by assumption, see Eq.~\eqref{assumptionAE1}. Finally, by writing explicitly the operator $\hat{S}^\dagger\hat{S}$ in the basis $\{\vert a\rangle,\vert b \rangle\}$, i.e., 
\begin{equation}
    \hat{S}^\dagger\hat{S}=\sum_{j,k=a,b}\frac{\Omega_{j}^*\Omega_{k}}{4}\vert j\rangle\langle k\vert,
\end{equation}
we can write the reduced dynamics of the system in the compact form
\begin{multline}
    \dot{\rho}_S(t) =
    \\-\frac{i}{\hbar}\left[\hat{H}_S-\sum_{j,k=a,b}\frac{\Omega_{j}^*\Omega_{k}}{4\Delta}\vert j\rangle\langle k\vert,\rho_S\right].
\end{multline}
In other words, the far-detuned level $\vert e \rangle$ effectively modifies the Hamiltonian of the states $\vert a \rangle$
and $\vert b\rangle$ to 
\begin{multline}
    \hat{H}_S'/\hbar = \left(-\frac{\delta}{2}-\frac{\vert\Omega_{a}\vert^2}{4\Delta}\right)\vert a\rangle\langle a\vert
    +
    \\
    +\left(\frac{\delta}{2}-\frac{\vert\Omega_{b}\vert^2}{4\Delta}\right)\vert b\rangle\langle b\vert
    +
    \\-
    \left[\frac{\Omega_{a}^*\Omega_{b}}{4\Delta}\vert a\rangle\langle b\vert + \text{H.c.}\right]
\end{multline}
which is the same result obtained in Ref.~\cite{BrionJPhysA2007}. We emphasize that this result could also have been obtained by second-order quasi-degenerate perturbation theory. The result of this section, however, evidences the power and applicability range of projection operator techniques beyond describing just bath-induced dissipation.

\section{Conclusions}\label{SecConclusions}

With this tutorial on projection operator techniques, we hope to have provided a further insight on how to develop effective dynamics in open quantum systems, especially for young scientists. On the one hand we have defined what common approaches such as adiabatic elimination and Born-Markov master equation mean in the context of perturbative expansions of the exact, global Nakajima-Zwanzig equation. We hope this will provide the reader with a deeper understanding of the origin of these approaches as well as their deep relationship (as discussed in the text, sometimes they are equivalent). On the other hand, beyond a deeper understanding of these methods, our tutorial provides a toolbox to analyze systems and regimes beyond conventional approaches, either via including further orders in the perturbative expansions or via devising new expansions suited to specific setups.
Although we consider that the contents of this tutorial are more than sufficient for most applied quantum theorists, they are just the tip of the iceberg in the world of open quantum systems. It is our hope that young scientists will find in this tutorial an open door to deeper explorations of this rich area of physics.

\section{Acknowledgments}

The author acknowledges Saurabh Gupta, Katja Kustura, Andreu Riera-Campeny, Oriol Romero-Isart, Oriol Rubies-Bigorda, and Cosimo Rusconi for their support, their revisions and improvements of the final manuscript, and their encouragement to make this tutorial publicly available.

\appendix

\section{A deeper insight: bath correlators and quantum regression theorem}\label{appendix}

This Appendix provides a deeper exploration of the derivation of master equations with decaying reservoirs (i.e., the particular cases of Secs.~\ref{SecBrownian} and \ref{SecBornMarkov}), beyond the ``practical toolbox'' approach of the main text. Specifically we will consider a coherent system-bath interaction and show how, within the weak system-bath coupling approximation, all the effects of the bath on the effective system dynamics is determined by the bath two-time correlation functions. We will also show how to compute these correlators using the quantum regression theorem and provide a simple example. The results in this appendix provide tools to generalize the techniques in this tutorial to more general systems.

We consider the same system, bath, and interaction Liouvillians discussed in Sec.~\ref{SecCoherentInteraction}, i.e., Liouvillians that in the interaction picture read 
\begin{equation}
    \Lv_S^{(i)} (*) =  \mathcal{D}_S^{(i)}(t)[(*)],
\end{equation}
\begin{equation}
    \Lv_B^{(i)} (*) = \mathcal{D}_B^{(i)}[(*)],
\end{equation}
\begin{multline}\label{LintMEQ3}
    \Lv_{\rm Int}^{(i)}(t) (*) = \\-\frac{i}{\hbar}\left[\hbar\sum_\alpha\hat{S}^{(i)}_\alpha(t)\otimes\hat{B}^{(i)}_\alpha(t),(*)\right],
\end{multline}
which have been redefined so that the bath operators $\hat{B}^{(i)}_\alpha(t)$ have zero expectation value. We can thus directly describe this system in the weak system-bath coupling approximation via Eq.~\eqref{NZMarkov22},
\begin{multline}\label{NakajimaAppendix}
    \dot{\rho}_S^{(i)}(t) \approx\mathcal{D}_S^{(i)}(t)[\rho_S(t)] + \text{Tr}_B  \Lv_{\rm Int}^{(i)}(t)\\
    \int_0^t d\tau e^{\mathcal{D}_B^{(i)}\tau} \mathcal{S}(t,\tau)
\Lv_{\rm Int}^{(i)}(t-\tau) v(t-\tau).
\end{multline}
with a dissipative system propagator given by
\begin{equation}
   \mathcal{S}(t,\tau) \equiv \mathcal{T}_+\left[\exp\int_0^\tau dt'\mathcal{D}_S^{(i)}(t-t')\right].
\end{equation}

Our goal is to derive the explicit expression for the system effective dynamical equation for the general interaction Liouvillian Eq.~\eqref{LintMEQ3}. To do so, we introduce Eq.~\eqref{LintMEQ3} into Eq.~\eqref{NakajimaAppendix} and explicitly trace out the bath degrees of freedom, obtaining
\begin{multline}\label{chorizo1}
    \dot{\rho}_S^{(i)}(t) =\mathcal{D}_S^{(i)}(t)[\rho_S(t)] - \int_0^td\tau \sum_{\alpha\beta} \Big[\hat{S}_\beta^{(i)}(t),
    \\
    f_{\hat{B}_\alpha\mathbb{1}\hat{B}_\beta}(t,\tau)
\mathcal{S}(t,\tau)\hat{S}_\alpha^{(i)}(t-\tau)\rho_S^{(i)}(t-\tau)
\\
-\mathcal{S}(t,\tau)\rho_S^{(i)}(t-\tau)\hat{S}_\alpha^{(i)}(t-\tau)
f_{\hat{B}_\alpha^\dagger\hat{B}_\beta^\dagger\mathbb{1}}(t,\tau)
\Big].
\end{multline}
Here, we have assumed $\rho_B^{(i)}(t)$ is time independent for simplicity and we define the functions
\begin{multline}
    f_{\hat{O}_\alpha\hat{O}_\beta\hat{O}_\gamma}(t,\tau) \equiv 
    \\
    \text{Tr}_B\left[\hat{O}^{(i)}_\beta (t)e^{\mathcal{D}_B^{(i)}\tau}\left(\hat{O}_\gamma^{(i)}(t-\tau)\rho_B^{(i)}\hat{O}^{(i)}_\alpha(t-\tau) \right)\right]
\end{multline}
with arbitrary bath operators $\hat{O}_\alpha \in \{\hat{B}_\alpha\}$. These functions of the bath are core to open quantum systems as they encode the full bath-induced effective dynamics of the system. Importantly, these functions are equal to the two-time correlation functions on the bath steady-state 
\begin{equation}\label{QRT}
    f_{\hat{O}_\alpha\hat{O}_\beta\hat{O}_\gamma}(t,\tau) = \langle 
    \hat{O}_\alpha(t)
    \hat{O}_\beta(t+\tau)
    \hat{O}_\gamma (t)\rangle_{\rm ss}.
\end{equation}
This identity is also known as the quantum regression theorem, and in the spirit of this practical tutorial we will not explore it in depth here. Curious readers can easily find proofs and deep discussions of the quantum regression theorem in the literature~\cite{CarmichaelBook,NavarreteQOCourse,GardinerZollerQNoise,BreuerPetruccione}.

The two-time correlator on the right-hand side of Eq.~\eqref{QRT} is defined for a bath that is uncoupled from the system, and the operators are in the Heisenberg picture. If the bath Liouvillian contains dissipation, as we are assuming here, then the operators inside the correlator are only well-defined in an extended Hilbert space not only containing the bath degrees of freedom, but also explicitly including the degrees of freedom of the additional reservoir that, when traced out, results in the dissipative behaviour of $\Lv_B$. Specifically, these Heisenberg operators must be defined as $\hat{O}_\alpha(t) \equiv \exp[it(\hat{H}_B + \hat{H}_C + \hat{V}_{BC})/\hbar] \hat{O}_\alpha \exp[-it(\hat{H}_B + \hat{H}_C + \hat{V}_{BC})/\hbar]$, where $\hat{O}_\alpha$ is the bath operator in the Schr\"odinger picture, $\hat{H}_B$ is the bath Hamiltonian, $\hat{H}_C$ is the Hamiltonian of the extra reservoir, and $\hat{V}_{BC}$ is their interaction (see an example in Sec.~\ref{exampleAppB}). Naturally this definition makes it complicated to compute the bath correlators if one only has knowledge of the traced-out bath Liouvillian $\Lv_B$, i.e., if the explicit forms of $\hat{H}_C$ and $V_{BC}$ are not known. Fortunately the second part of the quantum regression formula provides a route to perform this computation, as we will see below.

Once the two-time correlators have been defined, we can compactly cast the reduced dynamics Eq.~\eqref{NakajimaAppendix} as
\begin{multline}\label{chorizo1}
    \dot{\rho}_S^{(i)}(t) =\mathcal{D}_S^{(i)}(t)[\rho_S(t)] - \int_0^td\tau \sum_{\alpha\beta} \Big[\hat{S}_\beta^{(i)}(t),
    \\
C_{\hat{B}_\beta\hat{B}_\alpha}(\tau)\hat{A}_\alpha(t-\tau)-\hat{\overline{A}}_\alpha(t-\tau)C^*_{\hat{B}_\beta\hat{B}_\alpha}(\tau)\Big].
\end{multline}
Here we have defined the following correlation functions
\begin{equation}\label{corrgeneral}
    C_{\hat{O}_\alpha\hat{O}_\beta} (\tau)\equiv \langle\hat{O}_\alpha(t+\tau)\hat{O}_\beta\rangle_{\rm ss},
\end{equation}
whose dependence on only the argument difference $\tau$ stems from the properties of the steady state~\cite{CarmichaelBook}. We have also defined the following system operators,
\begin{equation}\label{Aop1}
    \hat{A}_\alpha(t-\tau) \equiv \mathcal{S}(t,\tau)\hat{S}_\alpha^{(i)}(t-\tau)\rho_S^{(i)}(t-\tau),
\end{equation}
\begin{equation}\label{Aop2}
    \hat{\overline{A}}_\alpha(t-\tau) \equiv \mathcal{S}(t,\tau)\rho_S^{(i)}(t-\tau)\hat{S}_\alpha^{(i)}(t-\tau).
\end{equation}
Note that Eq.~\eqref{chorizo1} is already a reduced equation for the system degrees of freedom, i.e., the bath degrees of freedom have been fully traced out. All the bath-induced dynamics is thus fully determined by the two-time bath correlation functions Eq.~\eqref{corrgeneral}. This is a core result of open quantum systems that only relies on the assumption of weak system-bath coupling. In the next two sections we particularize to the case where the bath correlators decay on a very fast timescale (Markov approximation) and show how to compute the two-time correlation functions.

\subsection{Markov approximation and linear coupling}

Equation~\eqref{chorizo1} is still non-local in time. This non-locality is also called non-Markovianity and makes solving the effective dynamics an involved task. To simplify it it is very common, as shown in the particular cases of Secs.~\ref{SecBrownian} and \ref{SecBornMarkov}, to undertake the further assumption that the bath degrees of freedom equilibrate to the steady state $\rho_B$ on a much faster timescale than the remaining timescales of the system~\cite{BreuerPetruccione}. This Markov approximation, which in the main text we have expressed in terms of eigenvalues of the bath Liouvillian (see e.g. Sec.~\ref{SecBrownian}), can in this general picture be more rigorously stated. Specifically, we say that a system is Markovian if the two-time correlation functions $C_{\hat{O}_\alpha,\hat{O}_\beta} (\tau)$ of the bath operators appearing in the interaction Liouvillian ($\{\hat{O}_\alpha\hat{O}_\beta\}\in\{\hat{B}_\alpha\}$) decay, as a function of $\tau$, much faster than the timescales associated to the system evolution, specifically much faster than the evolution of the operators $\{\hat{A}_\alpha(t-\tau),\hat{\overline{A}}_\alpha(t-\tau)\}$ with the variable $\tau$. Note that by the definition of these operators (Eqs.~\eqref{Aop1}-\eqref{Aop2}), this statement includes (i) the timescale at which the system reduced desity matrix $\rho_S^{(i)}(t)$ evolves in the interaction picture (typically given by the system-bath interaction rates) and (ii) the timescale at which the dissipative propagator $\mathcal{S}(t,\tau)$ evolves in the interaction picture (typically given by the system dissipation rates). In the case of the Brownian motion example of Sec.~\ref{SecBrownian} it also includes (iii) the timescale at which system operators $\hat{S}_\alpha^{(i)}(t)$ evolve in the interaction picture (typically the system eigenfrequencies). Here, however, we will focus on the more common Born-Markov paradigm of Sec.~\ref{SecBornMarkov}, where the system and bath coherent timescales are allowed to be faster than the decay time of the bath correlators.

If the system is Markovian according to the above definition, we can undertake the Markov approximation, which consists in
\begin{enumerate}
    \item Extending the upper limit of the integration to infinity as times larger than the correlator decay time do not contribute to the integral.
    \item Approximating
    \begin{equation}\label{Markov1}
        \mathcal{S}(t,\tau)\approx \mathcal{S}(t,0) = \mathcal{I}
    \end{equation}
    and
    \begin{equation}\label{Markov2}
        \rho_S^{(i)}(t-\tau) \approx \rho_S^{(i)}(t).
    \end{equation}
\end{enumerate}

Let us simplify Eq.~\eqref{chorizo1} under the Markov approximation. We assume the system Hamiltonian to be time-independent in the Schr\"odinger picture. Although not a necessity, for simplicity we also assume that all the original Liouvillians are quadratic, i.e., they contain at most up to quadratic combinations of system and bath ladder operators. This situation represents a system and a bath that are linearly coupled, which is a common scenario in open quantum systems~\cite{weiss2012quantum,CarmichaelBook,BreuerPetruccione}. For the interaction Liouvillian specifically, and assuming that none of the operators $\{\hat{S}_\alpha,\hat{B}_\alpha\}$ is the identity, linear coupling implies that these operators are linear combinations of creation and annihilation operators of their respective Hamiltonians, i.e., for the system we can write in the Schr\"odinger picture
\begin{equation}\label{Sintermsofas}
    \hat{S}_\alpha = \sum_l K_{\alpha l}\hat{v}_l  + \text{H.c.},
\end{equation}
with $\overleftrightarrow{K}$ an arbitrary matrix and $\hat{\mathbf{v}}\equiv[\hat{v}_1,...\hat{v}_N]$ a vector containing all the annihilation operators of the system Hamiltonian in the Schr\"odinger picture, i.e., $\left[\hat{H}_S,\hat{v}_l\right]=-i\Omega_l\hat{v}_l$ with $\Omega_l$ the system oscillation frequencies. Note that the statistics of the operators $\hat{v}_l$ are left unspecified. Note also that the above illustrative representation in terms of a finite amount of system modes remains valid in the case of infinite system degrees of freedom, $N\to \infty$, and in the case of continuous-mode systems where the mode index of the creation operators is a continuous index instead of a discrete one. 

By definition, the system operators can be written in the interaction picture as
\begin{equation}\label{SintermsofasIP}
    \hat{S}_\alpha^{(i)} = \sum_l K_{\alpha l}\hat{v}_le^{-i\Omega_l t}  + \text{H.c.}
\end{equation}
Using this identity and the Markov approximation Eqs.~\eqref{Markov1}-\eqref{Markov2} we can derive time-local approximations for the operators in Eqs.~\eqref{Aop1}-\eqref{Aop2}:
\begin{multline}
    \hat{A}_\alpha(t-
    \tau) \approx \\\sum_l\left(K_{\alpha l}\hat{v}_l^{(i)}(t)e^{i\Omega_l\tau}+\text{H.c.}\right)\rho_S^{(i)}(t),
\end{multline}
\begin{multline}
    \hat{\overline{A}}_\alpha(t-
    \tau) \approx \\\rho_S^{(i)}(t)\sum_l\left(K_{\alpha l}\hat{v}_l^{(i)}(t)e^{i\Omega_l\tau}+\text{H.c.}\right).
\end{multline}
Introducing these identities in Eq.~\eqref{chorizo1} and transforming back to the Schr\"odinger picture we obtain
\begin{multline}\label{Redfield}
    \dot{\rho}_S = \mathcal{L}_S\rho_S-\pi\sum_{\alpha\beta}\sum_{nm}\Big[
    K_{\beta n}\hat{v}_n + \text{H.c.}, 
    \\
    K_{\alpha l}\left(\hat{v}_l\rho_S S_{\hat{B}_\beta\hat{B}_\alpha}(\Omega_l) - \rho_s\hat{v}_lS^*_{\hat{B}_\beta\hat{B}_\alpha}(-\Omega_l)\right)
    \\
    +
    K_{\alpha l}^*\left(
    \hat{v}_l^\dagger\rho_S S_{\hat{B}_\beta\hat{B}_\alpha}(-\Omega_l) - \rho_s\hat{v}_l^\dagger S^*_{\hat{B}_\beta\hat{B}_\alpha}(\Omega_l)
    \right)
    \Big]
\end{multline}
where we define the bath power spectral densities
\begin{equation}\label{powerspectrum}
    S_{\hat{O}_\alpha\hat{O}_\beta}(\omega) \equiv \frac{1}{\pi}\int_0^\infty d\tau C_{\hat{O}_\alpha\hat{O}_\beta} (\tau) e^{i\omega \tau}.
\end{equation}

Equation \eqref{Redfield} is the effective system equation obtained within the Born-Markov approximations. It is called the Redfield or Bloch-Redfield equation~\cite{REDFIELD19651,RedfieldEM,BreuerPetruccione}. However, it is not always a satisfactory equation in this form, as the Redfield equation is known to not guarantee positivity of the density matrix~\cite{BreuerPetruccione}. To obtain a fully quantum-consistent master equation, the last step is to perform a rotating wave approximation, also known in this context as a secular approximation. This approximation assumes that the system frequencies $\Omega_l$ are non-degenerate and very different from each other, specifically~\cite{Fleming_2010}
\begin{equation}
    \vert \Omega_n + \Omega_l\vert \gg \vert \Omega_n-\Omega_l \vert \gg \Gamma_n^\pm
\end{equation}
for $n\ne l$, where $\Gamma_n^\pm$ are the system dissipation rates to be defined below. Under this approximation we can neglect all the non-diagonal elements $n\ne m$ in the sum, as well as the diagonal elements containing two creation ($\sim\hat{v}_n^{\dagger2}$) or two annihilation ($\sim\hat{v}_n^2$) operators from the same mode, obtaining
\begin{multline}\label{Lindblad0}
    \dot{\rho}_S = \mathcal{L}_S\rho_S
    \\
    -\pi\sum_{n}\left(G_{nn}\left[\hat{v}_n,\hat{v}_n^\dagger\rho_S\right]
    -
G_{nn}^*\left[\hat{v}_n^\dagger,\rho_S\hat{v}_n\right]\right)
\\
 +\pi\sum_{n}\left(Q_{nn}\left[\hat{v}_n,\rho_S\hat{v}_n^\dagger\right]
    -
Q_{nn}^*\left[\hat{v}_n^\dagger,\hat{v}_n\rho_S\right]\right),
\end{multline}
with rates
\begin{equation}
    G_{nn} \equiv \sum_{\alpha\beta}K_{\beta n}K^*_{\alpha n}S_{\hat{B}_\beta\hat{B}_\alpha}(-\Omega_n),
\end{equation}
and
\begin{equation}
   Q_{nn} \equiv \sum_{\alpha\beta}K_{\beta n}K^*_{\alpha n}S^*_{\hat{B}_\beta\hat{B}_\alpha}(\Omega_n).
\end{equation}
Writing these rates in terms of real and imaginary parts, i.e., $G_{nn}\equiv \Gamma_n^- + i \Delta_n^-$ and $Q_{nn} \equiv \Gamma_n^+ + i\Delta_n^+$, we can cast the master equation in Lindblad form (also called Lindblad-Gorini-Kossakowski-Sudarshan form)~\cite{Lindblad1,Lindblad2GoriniKossakowskiSudarshan},
\begin{multline}\label{Lindblad1}
    \dot{\rho}_S = \mathcal{L}_S\rho_S -\frac{i\pi}{\hbar}\left[\Delta_n^+\hat{v}^\dagger_n\hat{v}_n -\Delta_n^-\hat{v}_n\hat{v}^\dagger_n,\rho_S\right]
    \\
    +2\pi\sum_n \Gamma_n^-\left(\hat{v}_n^\dagger\rho_S\hat{v}_n-\frac{1}{2}\{\hat{v}_n\hat{v}_n^\dagger,\rho_S\}\right)
    \\
    +2\pi\sum_n \Gamma_n^+\left(\hat{v}_n\rho_S\hat{v}_n^\dagger-\frac{1}{2}\{\hat{v}_n^\dagger\hat{v}_n,\rho_S\}\right).
\end{multline}
The above equation completes the derivation of a general master equation and offers a deep insight into what open quantum system approaches imply, namely to characterize the bath -- regarding its interaction with the system -- in terms of only the power spectral densities evaluated at the system eigenfrequencies. Examples of these power spectral densities are given in the main text, for instance in Eq.~\eqref{PSDprov}. This dependence on the power spectrum can be extended to give a simple interpretation of the examples in the main text and of the behavior of many open systems. For instance, cavity cooling of a mechanical resonator (Sec.~\ref{exampleAEQO}) is enabled by modifying the optical density of states, that is, the power spectral density of the optical bath, so that $S(\Omega) \gg S(-\Omega)$. This increases the rate of the cooling dissipator with respect to the rate of the heating dissipator, resulting in net cooling. Understanding the core dependence of the master equation rates on bath power spectra enables to figure out novel schemes to tune them, e.g., cooling a mechanical resonator by using squeezed light to modify the power spectral densities~\cite{Teufelcooling}.
As a last remark, note that the appearance of only two-operator correlators for the bath (and not higher-order) is a consequence of our weak coupling approximation of the Nakajima-Zwanzig equation Eq.~\eqref{NakajimasimplifiedWC}, which is second order in the interaction Liouvillian. 

\subsection{Computing correlators}

In this short section we provide a short recipe for the computation of two-operator correlators $C_{\hat{O}_\alpha\hat{O}_\beta}(\tau)$ (Eq.~\eqref{corrgeneral}) appearing in the weak-coupling master equations. In the spirit of this tutorial we will focus only on providing a practical recipe. We address the curious reader to the well-known proofs found in the literature (two particularly didactic derivations of the results here can be found in Ref.~\cite{CarmichaelBook,NavarreteQOCourse}). 

We aim at calculating the correlation functions of the bath operators $\hat{B}_\alpha$ assuming the bath is uncoupled from the system but coupled to an external reservoir ``C''. In this scenario the dynamics of the bath ``B'' is governed by a non-conservative Liouvillian $\Lv_B$. The correlations $C_{\hat{O}_\alpha\hat{O}_\beta}(\tau)$ can be computed in two ways. The first option (see example~\ref{exampleAppB} below) is to do it by ``brute force'', namely, by coming up with a specific Hamiltonian for the external reservoir, $\hat{H}_C$, as well as a specific interaction Hamiltonian $\hat{V}_{BC}$, and computing the expectation values of the Heisenberg operators. In this route the form of $\hat{H}_C$ and $\hat{V}_{BC}$ have to be chosen such that after tracing out the reservoir ``C'' degrees of freedom one recovers the same dynamics for the bath ``B'' that are generated by the original Liouvillian $\Lv_B$. 

The second approach (see example~\ref{exampleAppA} below) is usually faster and more convenient and relies on the quantum regression theorem. It starts by assuming a complete set of bath operators $\hat{\mathbf{b}}\equiv[\hat{b}_1,...,\hat{b}_N]$ fulfilling for every bath operator $\hat{O}$ the following identity:
\begin{equation}\label{auxQRT1}
   \text{Tr}_B[\hat{b}_l(\mathcal{L}_B\hat{O})] = \sum_k M_{lk} \text{Tr}_B[\hat{b}_l\hat{O}],
\end{equation}
with $\overleftrightarrow{M}$ an arbitrary matrix.
Such complete set of operators always exists and they are typically easy to identify. Specifically, for quadratic bosonic systems these correspond to all creation and annihilation operators (see example below), whereas for spin systems they correspond to the Pauli matrices plus the identity matrix. A consequence of Eq.~\eqref{auxQRT1} is that we can write the time evolution of the operators in the set as a closed linear system,
\begin{equation}
    \frac{d}{dt}\langle \hat{\mathbf{b}}\rangle = \overleftrightarrow{M}\langle \hat{\mathbf{b}}\rangle.
\end{equation}
The core identity provided by the quantum regression formula establishes that two-time correlators involving one of these operators obeys the same dynamical equation in the delay variable $\tau$, that is, for two bath operators $\hat{O}_1$ and $\hat{O}_2$ we can write~\cite{CarmichaelBook,GardinerZollerQNoise,BreuerPetruccione}
\begin{multline}
    \frac{d}{d\tau}\langle \hat{O}_1(t)\hat{\mathbf{b}}(t+\tau)\hat{O}_2(t)\rangle =\\ \overleftrightarrow{M} \langle\hat{O}_1(t)\hat{\mathbf{b}}(t+\tau)\hat{O}_2(t)\rangle.
\end{multline}
This equation can be immediately solved formally to obtain the correlators in the steady state,
\begin{equation}
    \langle \hat{O}_1(t)\hat{\mathbf{b}}(t+\tau)\hat{O}_2(t)\rangle_{\rm ss} = e^{\overleftrightarrow{M}\tau}\langle \hat{O}_1\hat{\mathbf{b}}\hat{O}_2\rangle_{\rm ss},
\end{equation}
or, particularizing to a single operator,
\begin{multline}\label{QRTapp}
    \langle \hat{O}_1(t)\hat{b}_i(t+\tau) \hat{O}_2(t)\rangle=\\\mathbf{e}_ie^{\overleftrightarrow{M}\tau}\langle \hat{O}_1(t)\hat{\mathbf{b}}\hat{O}_2\rangle_{\rm ss},
\end{multline}
with $\mathbf{e}_i$ the $N-$dimensional unit vector along direction $i$.
These expressions allow to easily compute the correlators of any operator that can be expressed in terms of the elements $\hat{b}_i$. With them one can also compute the power spectral densities and the resulting master equation rates.

\subsubsection{Example: single harmonic oscillator}\label{exampleAppA}

As a simple example let us consider the bath ``B'' of example~\ref{brownianexample} in the main text, i.e., a bath formed by a single harmonic oscillator with frequency $\Omega$ decaying to a zero-temperature reservoir at rate $\gamma$. The Liouvillian is given by
\begin{equation}\label{Effdynforb}
    \Lv_B(*) = -i\Omega\left[\hat{b}^\dagger\hat{b},(*)\right] + \gamma \mathfrak{D}_{\hat{b}}[(*)].
\end{equation}
This is a quadratic bosonic system and hence the complete set of operators is given by $\hat{\mathbf{b}}\equiv [\hat{b},\hat{b}^\dagger]\equiv[\hat{b}_1,\hat{b}_2]$. Indeed, by computing the time evolution $(d/dt)\langle \hat{b}\rangle = \text{Tr}[\dot{\rho}_b\hat{b}] = \text{Tr}[ (\Lv_B\rho_b)\hat{b}]$ with $\rho_b$ an arbitrary density matrix of B, we easily find
\begin{multline}
    \frac{d}{dt}\hat{\mathbf{v}}=\frac{d}{dt}\left[\begin{array}{c}
         \langle \hat{b}\rangle  \\
         \langle \hat{b}^\dagger\rangle 
    \end{array}
    \right] = \\\left[
    \begin{array}{cc}
        -i\Omega-\gamma/2 & 0 \\
        0 & i\Omega-\gamma/2
    \end{array}
    \right]\left[\begin{array}{c}
         \langle \hat{b}\rangle  \\
         \langle \hat{b}^\dagger\rangle 
    \end{array}
    \right]
    \\
    \equiv\overleftrightarrow{M} \hat{\mathbf{v}}
\end{multline}

We aim at computing correlators of the bath operators appearing in the system-bath interaction Liouvillian, $\hat{B}_\alpha$. In the case of linear coupling these operators can be written as linear combinations of creation and annihilation operators (see the discussion in previous section),
\begin{equation}
    \hat{B}_\alpha = w_\alpha\hat{b}_1 + \text{H.c.}.
\end{equation}
Hence, we can write the target correlator, say $C_{\hat{B}_\alpha\hat{B}_\beta}(\tau)$, as
\begin{multline}
    C_{\hat{B}_\alpha\hat{B}_\beta}(\tau) = w_\alpha w_\beta C_{\hat{b}_1\hat{b}_1}(\tau)
    +
    w_\alpha w_\beta^* C_{\hat{b}_1\hat{b}_2}(\tau)
    \\
    +
    w_\alpha^* w_\beta C_{\hat{b}_2\hat{b}_1}(\tau)
    +
    w_\alpha^* w_\beta^* C_{\hat{b}_2\hat{b}_2}(\tau).
\end{multline}
The correlators in the right-hand side are easily computed using Eq.~\eqref{QRTapp},
\begin{equation}
    C_{\hat{b}_1\hat{b}_1}(\tau)=C_{\hat{b}_2\hat{b}_2}(\tau)=C_{\hat{b}_2\hat{b}_1}(\tau)=0,
\end{equation}
\begin{equation}\label{Corrnon0}
    C_{\hat{b}_1\hat{b}_2}(\tau) = e^{(-i\Omega-\kappa/2)\tau},
\end{equation}
where we have used the fact that in the steady state of $\Lv_B$ (the vacuum) we have $\langle \hat{b}^2\rangle = \langle \hat{b}^\dagger\hat{b}\rangle = 0$ and $\langle\hat{b}\hat{b}^\dagger\rangle=1$.
The original correlator thus reads
\begin{equation}
    C_{\hat{B}_\alpha\hat{B}_\beta}(\tau) =
    w_\alpha w_\beta^*e^{(-i\Omega-\kappa/2)\tau},
\end{equation}
and its corresponding power spectral density Eq.~\eqref{powerspectrum} reads
\begin{equation}
    S_{\hat{B}_\alpha\hat{B}_\beta}(\omega) = \frac{w_\alpha w_\beta^*}{\pi}\frac{1}{\kappa/2 +i(\Omega-\omega)}.
\end{equation}
The quantum regression formula in this section thus provides a fast practical way to compute master equation rates.

\subsubsection{Example: calculating correlators from Heisenberg operators}\label{exampleAppB}

Let us show, for the sake of illustration, how the same correlators can be calculated using the direct approach, namely enlarging the Hilbert space to include the reservoir inducing the decay of our bath. We consider the same harmonic oscillator as in the previous example, and we assume it is coupled to a continuum-mode reservoir of harmonic oscillators labelled by their frequency $\omega$ and with annihilation operators $\hat{c}_\omega$. The Hamiltonian is thus
\begin{equation}
    \hat{H} = \hat{H}_S + \hat{H}_B + \hat{V}_{BC},
\end{equation}
with 
\begin{equation}
    \hat{H}_S = \hbar\Omega\hat{b}^\dagger\hat{b},
\end{equation}
\begin{equation}
    \hat{H}_B = \hbar\int_0^\infty d\omega \omega \hat{c}^\dagger_\omega\hat{c}_\omega,
\end{equation}
\begin{equation}
    \hat{V}_{BC} = \sqrt{\frac{\gamma}{2\pi}}\int_0^\infty d\omega \left(\hat{b}\hat{c}^\dagger_\omega+\text{H.c.}\right).
\end{equation}
If we assume the reservoir ``C'' fulfills all the conditions to act as a Markovian bath and its steady state is the vacuum, then it can be shown that tracing it out using the techniques of Sec.~\ref{SecBornMarkov} recovers the effective dynamics Eq.~\eqref{Effdynforb} for the system ``B'' (except for a trivial frequency renormalization which could have been included a priori into $\hat{H}_S$ but which we neglect for simplicity). 
We aim at computing the correlator
\begin{multline}\label{correlatorbruteforce}
    C_{\hat{b}_1\hat{b}_2}(\tau) =C_{\hat{b}\hat{b}^\dagger} =\lim_{t\to \infty}\langle\hat{b}(t+\tau)\hat{b}^\dagger(t)\rangle
    \\
    = e^{-i\Omega \tau}
    \lim_{t\to \infty}\langle\hat{b}^{(i)}(t+\tau)\hat{b}^{\dagger(i)}(t)\rangle
\end{multline}
where $\hat{b}(t)$ are operators in the Heisenberg picture and in the last line we have transformed to the interaction picture for convenience. The remaining correlators in Example~\ref{exampleAppA} are computed in an analogous way.

We start by computing the dynamical equations for the interaction picture operators,
\begin{equation}\label{eomb}
    \frac{d}{dt}\hat{b}^{(i)}(t) = -i\sqrt{\frac{\gamma}{2\pi}}\int_0^\infty d\omega e^{-i\Delta t}\hat{c}_\omega^{(i)}(t),
\end{equation}
\begin{equation}\label{eomc}
    \frac{d}{dt}\hat{c}_\omega^{(i)}(t)=  -i\sqrt{\frac{\gamma}{2\pi}}\hat{b}^{(i)}(t)e^{i\Delta t},
\end{equation}
where we define $\Delta\equiv \omega-\Omega$. We now formally integrate Eq.~\eqref{eomc} between an initial time $t=0$ and current time $t$ and intoduce it into Eq.~\eqref{eomb}, casting it into the form of a Langevin equation,
\begin{multline}\label{langevin0}
    \frac{d}{dt}\hat{b}^{(i)}(t) = \hat{f}(t) \\-\frac{\gamma}{2\pi}\int_0^\infty d\omega \int_0^td\tau \hat{b}^{(i)}(\tau)e^{-i\Delta (t-\tau)},
\end{multline}
with a noise operator (sometimes also called input operator in input-output theory~\cite{GardinerZollerQNoise})
\begin{equation}
    \hat{f}(t)\equiv -i\sqrt{\frac{\gamma}{2\pi}}\int_0^\infty d\omega e^{-i\Delta t}\hat{c}_\omega^{(i)}(0).
\end{equation}
The noise operator has the properties of quantum white noise for a zero-temperature reservoir~\cite{GardinerZollerQNoise},
\begin{equation}
    \langle\hat{f}(t)\rangle_{ss,c} = 0,
\end{equation}
\begin{multline}
    \langle\hat{f}(t)\hat{f}(t')\rangle_{ss,c} = \langle\hat{f}^\dagger(t)\hat{f}^\dagger(t')\rangle_{ss,c}= 
    \\
    =\langle\hat{f}^\dagger(t)\hat{f}(t')\rangle_{ss,c} =0,
\end{multline}
\begin{equation}\label{langevincorrelator}
    \langle\hat{f}(t)\hat{f}^\dagger(t')\rangle_{ss,c} = \gamma\delta(t-t')
\end{equation}
where the subindex \textit{``ss,c''} indicates the steady state of the reservoir ``C'', i.e., its vacuum state.

Our next step is to perform a Markov approximation on Eq.~\eqref{langevin0}. This approximation takes into account that for sufficiencly large times $t$ the main contribution to the frequency integral will be from frequencies $\omega\approx\Omega$, provided that the operator $\hat{b}^{(i)}(t)$ evolves on a much slower timescale than $\omega_0$. Then one can approximate (see e.g.~\cite{CohenAtomPhotonInteractions,steck2007quantum} for details)
\begin{multline}
    \int_0^\infty d\omega \int_0^td\tau \hat{b}^{(i)}(\tau)e^{-i\Delta (t-\tau)}
    \\\approx \int_{-\infty}^\infty d\omega \int_0^td\tau \hat{b}^{(i)}(\tau)e^{-i\Delta (t-\tau)}
    \\
    =2\pi \int_0^td\tau \hat{b}^{(i)}(\tau) \delta(t-\tau) = \pi \hat{b}^{(i)}(t),
\end{multline}
so that the Langevin equation Eq.~\eqref{langevin0} takes the approximate form
\begin{multline}
    \frac{d}{dt}\hat{b}^{(i)}(t) = \hat{f}(t) \\-\frac{\gamma}{2}\hat{b}^{(i)}(t).
\end{multline}
The solution to this equation is
\begin{multline}
    \hat{b}^{(i)}(t) = \hat{b}^{(i)}(0)e^{-\gamma t/2} +\int_0^tdt'e^{-\gamma(t-t')/2}\hat{f}(t'),
\end{multline}
and in the long-time limit it loses the information about the initial value, i.e.,
\begin{equation}
    \lim_{t\to\infty}\hat{b}^{(i)}(t) = \lim_{t\to\infty}\int_0^tdt'e^{-\gamma(t-t')/2}\hat{f}(t').
\end{equation}

Using the above result we can explicitly compute the correlator Eq.~\eqref{correlatorbruteforce},
\begin{multline}
    C_{\hat{b}_1\hat{b}_2}(\tau) = e^{-i\Omega \tau}\lim_{t\to\infty}\int_0^tdt'e^{-\gamma(t+\tau-t')/2}
    \\ \times \int_0^tdt''e^{-\gamma(t-t'')/2}
    \langle\hat{f}(t')\hat{f}^\dagger(t'')\rangle_{\rm ss,c}
    =\\=
    \gamma e^{-i\Omega t}\lim_{t\to\infty}\int_0^tdt'e^{-\gamma(t-t'+\tau/2)} = e^{(-i\Omega-\gamma/2)\tau},
\end{multline}
where we have used Eq.~\eqref{langevincorrelator}. The above expression is the same result obtained in Eq.~\eqref{Corrnon0} using the much simpler approach based on the quantum regression formula.

%\medskip
 
%\printbibliography
\bibliographystyle{quantum}
\bibliography{bibliography}

\end{document}